\documentclass[manuscript]{aastex}
\slugcomment{Astrophys. J., in press}

\shorttitle{ }
\shortauthors{Pogorelov et al.}

\usepackage{epsfig}
\def\ssr{{Space Sci. Rev.}}
\def\jgr{{J.~Geophys.~Res.: Space Phys.}}
\def\aap{{Astron. Astrophys.}}
\def\apj{{Astrophys.~J.}}
\def\apjl{\apj}
\def\grl{{Geophys. Res. Lett.}}

\def\adsr{{Adv.~Space Res.}}

\def\f{\mathrm{f}}

\def\A{\mathrm{A}}
\usepackage{amssymb}
\begin{document}


\title{Three-dimensional Features of the Outer Heliosphere Due to Coupling between the Interstellar and Heliospheric Magnetic Field.~V.\\
The Bow Wave, Heliospheric Boundary Layer, Instabilities, and Magnetic Reconnection}

\author{N.~V.~Pogorelov,$^{1,2}$ J. Heerikhuisen$^{1,2}$, V.~Roytershteyn,$^3$ L.~F.~Burlaga,$^4$ D.~A.~Gurnett,$^5$  and W.~S.~Kurth$^5$}
\affil{
$^1$Department of Space Science, The University of Alabama in Huntsville, AL 35805, USA\\
$^2$Center for Space Plasma and Aeronomic Research, The University of Alabama in Huntsville, AL 35805, USA\\
$^3$Space Science Institute, Boulder, CO 80301, USA\\
$^4$NASA Goddard Space Flight Center, Greenbelt, MD 20771, USA\\
$^5$Department of Physics and Astronomy, The University of Iowa, Iowa City, IA 52242, USA}
\email{nikolai.pogorelov@uah.edu (corresponding author)}

\begin{abstract}
The heliosphere is formed due to interaction between the solar wind (SW) and local interstellar medium (LISM). The shape and position of the heliospheric boundary, the heliopause, in space depend on the parameters of interacting plasma flows. The interplay between the asymmetrizing effect of the interstellar magnetic field and charge exchange between ions and neutral atoms plays an important role in the SW--LISM interaction. By performing three-dimensional, MHD plasma / kinetic neutral atom simulations, we determine the width of the outer heliosheath -- the LISM plasma region affected by the presence of the heliosphere -- and analyze quantitatively the distributions in front of the heliopause. It is shown that charge exchange modifies the LISM plasma to such extent that the contribution of a shock transition to the total variation of plasma parameters becomes small even if the LISM velocity exceeds the fast magnetosonic speed in the unperturbed medium. By performing adaptive mesh refinement simulations, we show that a distinct boundary layer of decreased plasma density and enhanced magnetic field should be observed on the interstellar side of the heliopause. We show that this behavior is in agreement with the plasma oscillations of increasing frequency observed by the plasma wave instrument onboard \emph{Voyager }1. We also demonstrate that \emph{Voyager} observations in the inner heliosheath between the heliospheric termination shock and the heliopause are consistent with dissipation of the heliospheric magnetic field. The choice of LISM parameters in this analysis is based on the simulations that fit observations of energetic neutral atoms performed by \emph{IBEX}.
\end{abstract}


\section{Introduction}
\begin{figure*}[t]
\centering
\includegraphics[width=0.48\textwidth]{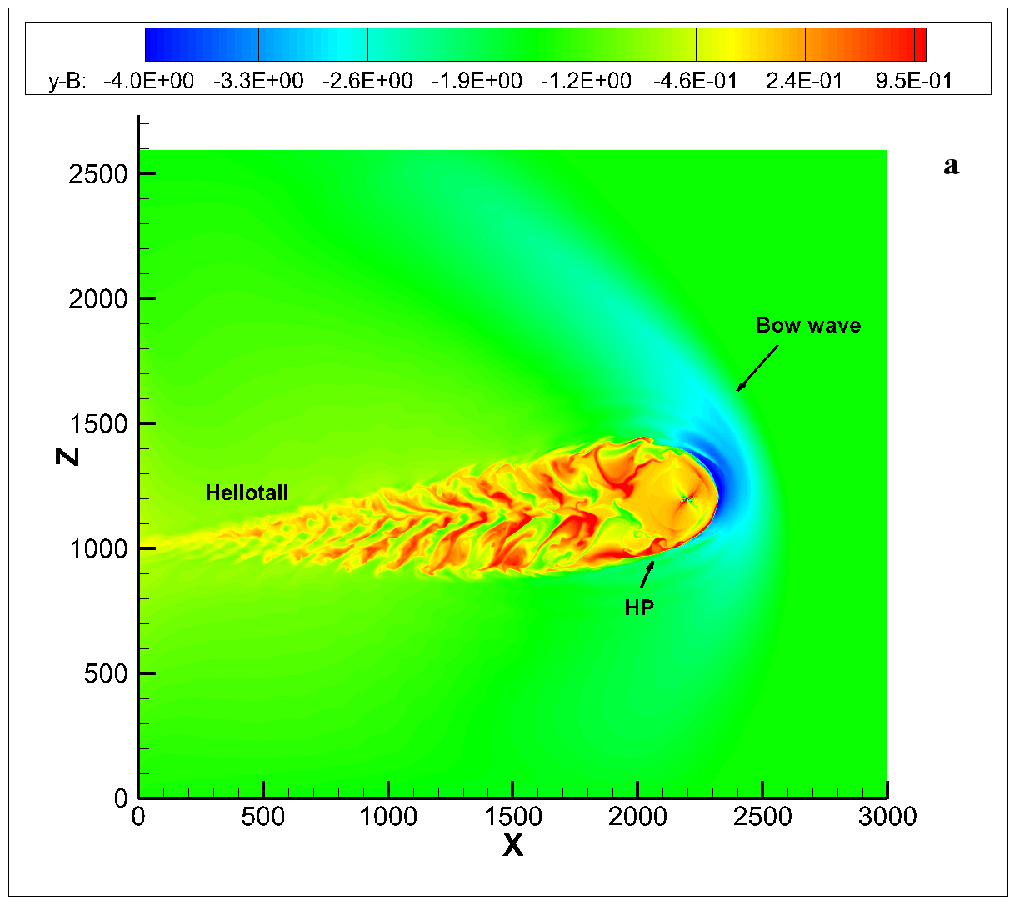}
\includegraphics[width=0.48\textwidth]{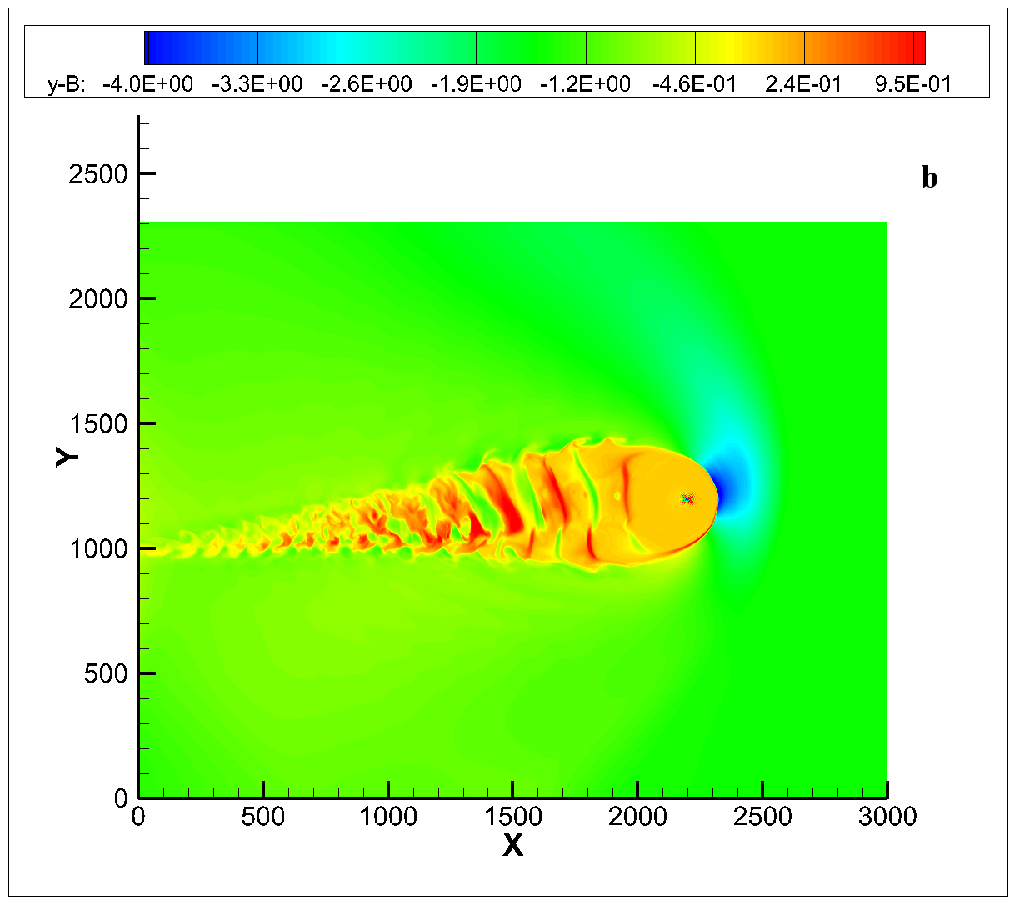}\\
\includegraphics[width=0.48\textwidth]{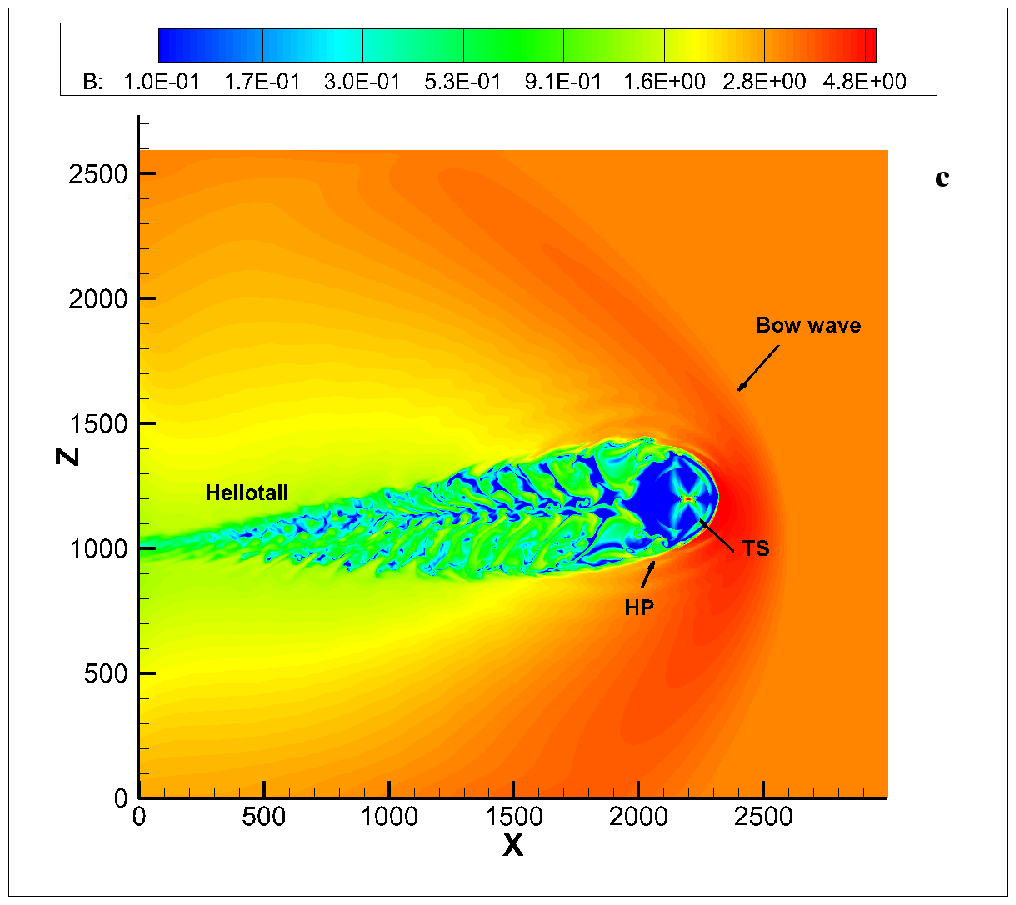}
\includegraphics[width=0.48\textwidth]{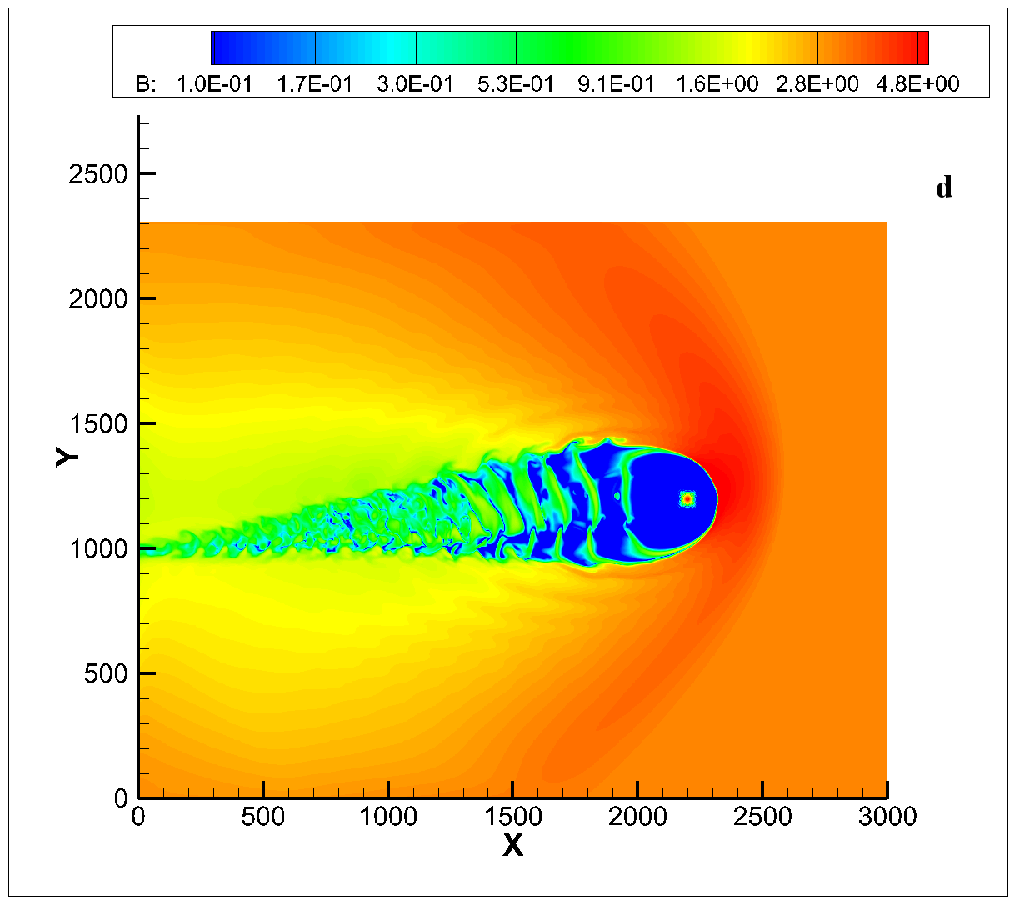}
\caption{SW--LISM interaction pattern in the presence of solar cycle effects.
The $y$-component (top panels) of the magnetic field vector and its magnitude (bottom panels) in the meridional (left panels) and equatorial (right panels) planes. Distances are given in AU and magnetic field in $\mu$G. One can see the TS, HP, and bow wave.
}
\label{fig1}
\end{figure*}

The interaction of the solar wind (SW) with the local interstellar medium (LISM) is essentially the combination of a blunt-body
and a supersonic jet flows. Head-on collision of the SW and LISM plasma flows creates a tangential discontinuity (the heliopause, HP), which
extends far into the wake region (see Fig.~\ref{fig1}). The SW flow in the direction parallel to the Sun's motion resembles a jet immersed into a medium with lower thermal pressure. The LISM plasma is decelerated at the HP, which may result, depending on the LISM parameters, in the formation of a so-called bow shock (BS). The SW flow, on the other hand, is decelerated due to its interaction with the HP, charge exchange with interstellar neutral atoms, and by the LISM counter-pressure in the heliotail region. Since the neutral hydrogen (H) density in the LISM is greater that the proton density, resonant charge exchange between ions and neutral ions plays a major role in the SW--LISM interactions \citep{1969Natur.223..936B,1977Holzer,1971NPhS..233...23W,1975Natur.254..202W}. In particular, the SW in the tail is decelerated and cooled down by charge exchange until the heliotail disappears at a few tens of thousands of AU. Because of the large mean free path, charge exchange and, in general, the transport of neutral atoms should be performed kinetically, by solving the Boltzmann equation. The first self-consistent simulation of this kind was performed by \citet{Bama93} in an axially-symmetric statement of the problem neglecting the effect of the heliospheric and interstellar magnetic fields (HMF and ISMF). This model was extended to time-dependent \citep {Izmod05b} and 3D flows \citep{Izmod05} much later. Another class
of models assume that neutral atoms can be treated as a fluid, or rather a set of fluids, each of them describing the flow of neutral atoms
born in thermodynamically different regions of the SW--LISM interaction. These are usually (i) the unperturbed LISM; (ii) the LISM region
substantially modified by the presence of the heliosphere; (iii) the region between the TS and HP; and (iv) the supersonic SW region
\citep{Pauls95,Zank96,Fahr-etal-2000,2004ApJ...604..700F,Pozaog06}. Such multi-fluid approaches are easily applicable to genuinely time-dependent problems \citep[see, e.g.,][]{Zank03,Sternal,Pogo09a,Pogo13b}, which are very expensive computationally when neutrals atoms are treated kinetically \citep{Izmod05b,Erik15b}.

Figure~\ref{fig1} shows a typical simulation result that takes into account solar cycle effects. The inner boundary conditions, corresponding to a nominal solar cycle with the radial velocity, ion density, and temperature in the fast and slow SW, are specified at the Earth orbit ($R=1$~AU).
%$V_\mathrm{Ef}=762$~$\mathrm{km} \,\mathrm{s}^{-1}$, $n_\mathrm{Ef}=2.4$~$\mathrm{cm}^{-3}$, $T_\mathrm{Ef}=2.45\times 10^5$~K, %$V_\mathrm{Es}=450$~$\mathrm{km} \,\mathrm{s}^{-1}$, $n_\mathrm{Es}=6.9$~$\mathrm{cm}^{-3}$, and $T_\mathrm{Es}=6.8\times 10^4$~K, %respectively.
%The radial component of the heliospheric magnetic field is $B_{R\mathrm{E}}= 35\ \mu$G.
It is assumed that the latitudinal extent of the slow wind varies with an 11-year period from $\theta=28^\circ$ at solar minima to $90^\circ$ at solar maxima. Additionally, the tilt between the Sun's magnetic and rotation axes varies from $\theta=8^\circ$ at solar minima to $90^\circ$ at solar maxima and flips to the opposite hemisphere at each maximum. This creates a sequence of regions possessing opposite HMF polarities in the heliotail.
%\textbf{The SW boundary conditions are from \citet{Borov14}. The quantities in the unperturbed LISM are from \citet{Jacob10}.}
%The LISM velocity and magnetic field magnitudes are $V_\infty=26.4$~$\mathrm{km} \,\mathrm{s}^{-1}$ and $B_\infty=3 \ \mu$G, respectively. The %LISM temperature is $T_\infty=8000$~K. The ion and neutral H number densities are $n_\infty=0.082$~$\mathrm{cm}^{-3}$ and %$n_{\mathrm{H}\infty}=0.172$~$\mathrm{cm}^{-3}$.
We perform all simulations in a so-called heliospheric coordinate system, where the $z$-axis is aligned with the Sun's rotation axis,
the $x$-axis belongs to the plane formed by the $z$-axis and $\mathbf{V}_\infty$, and directed upstream into the LISM. The $y$-axis completes the right coordinate system.
%In this coordinate system, the direction of $\mathbf{V}_\infty$ and $\mathbf{B}_\infty$ are aligned with the %vectors $(-0.996, 0, -0.089)$ and $(0.692, -0.477, 0.541)$, respectively.
The boundary conditions in the SW and LISM are taken from the existing simulation \citep{Borov14} and are for illustration purposes only. In summary, the heliosphere is characterized by the presence of a very long heliotail, which extends to distances exceeding 5,000~au, and is compressed approximately in the direction perpendicular to the $BV$-plane. The latter is defined by the LISM velocity and ISMF vectors, $\mathbf{V}_\infty$  and $\mathbf{B}_\infty$, in the unperturbed LISM. The width of the heliotail in the plane of its maximum flaring
(the $BV$-plane) decreases with distance from the Sun. This creates an illusion that the heliotail disappears when we look at the mutually perpendicular cross-sections shown in Fig.~\ref{fig1}. Three-dimensional pictures of the heliosphere can be found in \citet{Borov14}.
We will discuss the boundary conditions in the LISM in  Section~\ref{sec:bow}.

In principle, some sort of kinetic treatment of neutral atoms is preferred because the charge exchange mean free path is about 50--100~AU,
depending on the region of the heliosphere and the origin of H atoms. This is especially important for simulations aimed to provide input to calculations of energetic neutral atoms (ENAs) observed by the \emph{Interstellar Boundary Explorer} (\emph{IBEX}), see \citet{McComas17}
for a review of the mission results over the past 7 years. In particular, the secondary H atoms born in the SW are of importance
if we are interested in the distance to which the heliosphere may affect the LISM flow. It is known from theory and simulations
\citep{Gruntman,Bama93,Zank96} that secondary neutral atoms can travel far upwind where they may experience charge exchange and affect the LISM flow.
A number of the \textit{IBEX} ribbon models \citep{Jacob10,Chalov10,Isenberg,Giacalone} involve secondary neutral atoms.
Radio emission in a 2--3 kHz range observed by \textit{V1} also relies upon global shock waves propagating outward due to various solar events and ``the neutral SW'' (H atoms born inside the TS) as
a source of pickup ions (PUIs) that initiate a series of physical processes which ultimately result in the observed wave activity
\citep{Gurnett93,Gurnett06,Gurnett13,Gurnett15,Cairns02,Pogo08,Pogo09b,Mitchell}.
It is believed that the ring-beam instability of PUIs born in the outer heliosheath (OHS), i.e., in the region of the LISM affected by the presence of the heliosphere, resonantly accelerate ambient electrons by lower hybrid waves. These pre-accelerated electrons are further accelerated by transient shocks creating the foreshock electron beams, plasma waves, and radio emission.
Secondary neutral atoms are also important
to establish the geometrical size the OHS.
The speed and temperature of the unperturbed LISM can be derived from the properties of He atoms observed by such Earth-bound spacecraft as \textit{Ulysses} and \textit{IBEX} \citep{Witte,Bzowski,McComas15a}. It was shown that the pristine LISM flow is supersonic and one would expect a bow shock to be formed in front of the HP. However, the LISM is magnetized, so its flow may become subfast magnetosonic (its speed being less
than the fast magnetosonic speed), which will eliminate the fast-mode bow shock. In principle, slow-mode shocks may still exist
in front of the HP \citep{2004ApJ...604..700F,Pozaog06,Pogo11,Zieger} if the angle between $\mathbf{V}_\infty$ and  $\mathbf{B}_\infty$ is small.
In this paper, we will show that this is an unlikely scenario in the presence of charge exchange of LISM ions with secondary H atoms
because a fast-mode shock is not just disappearing when $B_\infty$ reaches some threshold value. It is eroding, its strength decreasing until
no shock is observed.
\begin{table*}[t]
\scriptsize
\label{tbl:1}
\begin{center}
\begin{tabular}{|c|c|c|c|c|c|c|c|c|}
\hline
Model& $n_\infty$, cm$^{-3}$ & $n_{\mathrm{H}\infty}$, cm$^{-3}$  & $B_\infty$, $\mu$G & $\mathbf{B}_\infty/ B_\infty$ & $V_\infty$, km s$^{-1}$ & $\mathbf{V}_\infty/ V_\infty$
& $T_\infty$, K\\
\hline
  1& 0.11   & 0.165  & 2 & $(0.806,-0.383,0.452)$ & 25.4 & $(-0.996,0,0.088)$ & 7500\\
 2 & 0.1   & 0.1595  & 2.5 & $(0.760,-0.426,0.491)$ & 25.4 &
 $(-0.996,0,0.088)$ & 7500 \\
  3& 0.095   & 0.157  & 2.75 & $(0.743,-0.441,0.504)$ & 25.4 &
  $(-0.996,0,0.088)$ & 7500 \\
  4& 0.09   & 0.154  & 3 & $(0.725,-0.455,0.517)$ & 25.4 &
  $(-0.996,0,0.088)$ & 7500\\
  5& 0.08   & 0.1495  & 3.5 & $(0.692,-0.480,0.539)$ & 25.4 & $(-0.996,0,0.088)$ & 7500\\
  6& 0.07   & 0.145  & 4 & $(0.664,-0.500,0.556)$ & 25.4 & $(-0.996,0,0.088)$ & 7500\\
\hline
\end{tabular}
\end{center}
\caption{Model description for our MHD plasma / kinetic neutral atoms simulation of the SW--LISM interaction.}
\end{table*}%

The objective of this paper is to investigate the structure of the LISM region perturbed by the presence of the heliosphere as a function
of LISM parameters. In particular, we determine the width of the LISM region perturbed by the heliosphere and the contribution of a shocked transition to the overall change of LISM properties across this region. In addition, we will consider some issues related to the formation of a boundary layer in the LISM plasma near the HP, development of instabilities and magnetic reconnection, and the HMF distribution in the presence of the heliospheric current sheet (HCS). Different problems require different models for their solution. For this reason, we use an MHD-kinetic model to investigate the bow shock behavior and the distribution of quantities in the LISM flowing around the heliopause.
On the other hand, a multi-fluid approach is more appropriate for modeling the HP instabilities.
Comparisons between MHD-kinetic and multi-fluid models are presented
by \citet{Alexashov05,Jacob06,Hans08} and \citet{Pogo09c}.
\begin{table}[t]
\scriptsize
\label{tbl:2}
\begin{center}
\begin{tabular}{|c|c|c|}
\hline
Model& $(\lambda_{V\infty}, \beta_{V\infty})\ (^\circ)$& $(\lambda_{B\infty}, \beta_{B\infty})\ (^\circ)$\\
\hline
  1& (255.7, 5.0)   &  (233.20, 29.86) \\
 2 & (255.7, 5.0)  & (229.61, 32.83)  \\
  3& (255.7, 5.0)  & (228.34, 33.81)  \\
  4&(255.7, 5.0)   & (226.99, 34.82)  \\
  5& (255.7, 5.0)   & (224.46, 36.61)  \\
  6& (255.7, 5.0)   & (222.31, 38.02)  \\
\hline
\end{tabular}
\end{center}
\caption{The directions of the unperturbed LISM velocity and ISMF vectors from Table~1 in the ecliptic J2000 coordinate system.}
\end{table}%

\section{Constraints on the LISM Properties}
As discussed in the Introduction, the LISM velocity vector and temperature can be derived from the He observations in the inner heliosphere.
The rest of quantities in the unperturbed LISM are not measured directly.
Following \citet{Erik16}, they can be chosen to satisfy a number of observational results:
\begin{enumerate}
\item
By analyzing the Ly-$\alpha$ backscattered emission in the \emph{Solar and Heliospheric Observatory} (\emph{SOHO}) solar win anisotropy (SWAN) experiment, \citet{Lallement05,Lallement10} discovered a deflection ($\sim 5^\circ$) of the neutral H atom flow in the inner
heliosphere from its original direction, $\mathbf{V}_\infty$. These two directions define a so-called hydrogen deflection plane (HDP).
MHD-plasma/kinetic-neutrals simulations \citep{Izmod05,Pogo08,Pogo09b,Katushkina} showed that the average deflection occurs predominantly parallel to the $BV$-plane.

\item
It is also possible to restrict LISM properties by fitting the \textit{IBEX} ribbon of enhanced ENA flux
\citep[see][where it was determined that the directions towards the ribbon strongly correlate with the lines of sight perpendicular to the
ISMF lines draping around the HP]{Nathan09}. As shown by \citet{Pogo10} and \citet{Jacob11}, the position of the
ENA ribbon strongly depends on rotation of the
$BV$-plane about the $\mathbf{V}_\infty$ vector. Kinetic ENA flux simulations of \citet{Jacob10,Jacob14,Jacob11,Erik15a,Erik15b,Erik16} reproduced the ribbon using the $BV$-plane consistent with the HDP. It is worth noticing here that the accuracy of \textit{SOHO} SWAN measurements allows substantial margins in the determination of the $BV$-plane \citep[see, e.g.,][]{Pogo07}. It is of interest from this viewpoint that the $BV$-plane from \citet{Erik16} lies in the middle of the range derived from the HDP analysis.
Astrophysical observations restricting the ISMF properties \citep{Frisch15} are also consistent with the above considerations.
New \textit{IBEX} results \citep{McComas17} also suggest secondary ENA's as accepted ribbon source.

\item
One-point-per-time, \emph{in situ} measurements performed in the LISM by \textit{V1}, also
provide restriction on the direction and strength of $\mathbf{B}_\infty$. E.g., numerical simulations of \citet{Pogo09b}
provided $\mathbf{B}_\infty\cdot\mathbf{R}=0$ ($\mathbf{R}$ is a unit vector in the radial direction) directions consistent with the \emph{IBEX} ribbon \citep{McComas09}.  The same choice of the LISM properties also reproduced the elevation and azimuthal angles in the ISMF beyond the HP \citep[see][]{Pogo13b,Borov14}. Simulations presented in \citet{Erik16} are also restricted by the HP position
and magnetic field angles observed by \textit{V1}. It is certainly possible \citep[see, e.g.,][]{Pogo15} to shift the HP position to 122 AU where the HP was crossed by \textit{V1} \citep{Gurnett13}. However, the TS position in the \textit{V1} direction becomes substantially smaller (by $\sim 20$~AU) than at the time of crossing. This should not be discouraging, since there is no information about the TS position after \textit{Voyagers}
crossed it. Numerical simulations based on \textit{Ulysses} observations \citep{Pogo13b} indeed show that the TS was moving inward between
2004 and 2010. A decrease in the SW ram pressure from one solar cycle to another could contribute to the HP shift inward by a few AU.
As shown in \citet{Malama} and \citet{Pogo16}, the IHS width also decreases when PUIs are treated as a separate plasma component.

\item
The H density at the TS derived from PUI measurements \citep{Bzowski09} can also be used to constrain the models.

\item
Modeled anisotropy in the 1--10 TeV galactic cosmic ray (GCR) flux and its comparison with multiple air shower observations
can also improve our knowledge of the LISM \citep{Nathan14,Ming14,Ming16}.
\end{enumerate}

\section{Bow Wave and Heliospheric Boundary Layer}
\label{sec:bow}
\citet{McComas12} and \citet{Zank13} investigated conditions to be satisfied for the bow shock to exist. However, a definitive answer to this question depends on the details of the global SW--LISM simulation and is not readily available from a direct analysis of the LISM properties far away from the HP. The LISM near the HP is a weakly collisional medium (the mean free path with respect to Coulomb collisions is about 1~AU, Baranov \& Ruderman, 2013), so the bow shock, if it exists, is rather well described by ideal MHD equations, which have the $t$-hyperbolic type. The source terms in these equations are due to charge exchange and therefore contain no delta-functions. As a result, the Hugoniot-type boundary conditions at a bow shock cannot be modified by such source terms. Thus, the only ``shock structure'' to be expected is of numerical origin. However, charge exchange modifies plasma and magnetic field both in front and behind the shock in a way unknown before the problem of the SW--LISM interaction is solved as a whole. It has been shown in the gas dynamic plasma ($\mathbf{B}=\mathbf{0}$)/ kinetic neutrals simulations of \citet{Izmod00}, the increase in the LISM neutral H density results in a weaker bow shock. This conclusion holds in the presence of ISMF.

\begin{figure*}[p]
\centering
\includegraphics[width=0.45\textwidth]{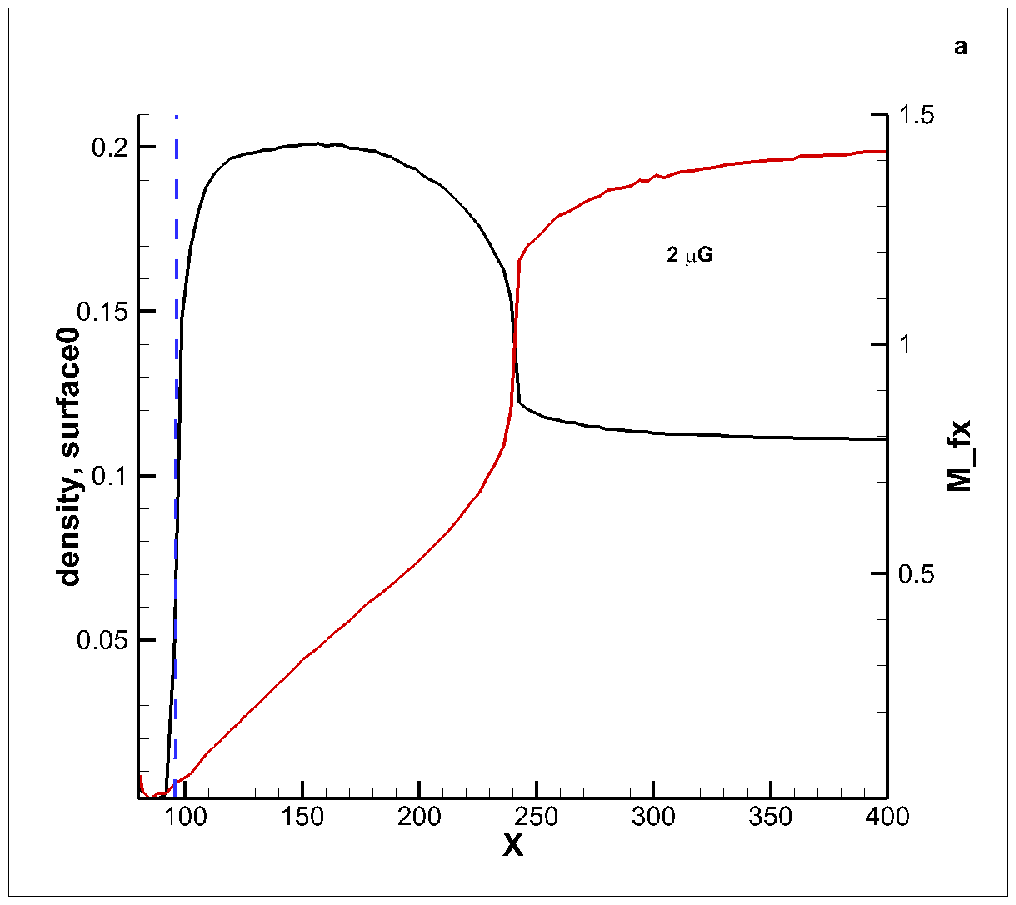}\hspace{5mm}
\includegraphics[width=0.45\textwidth]{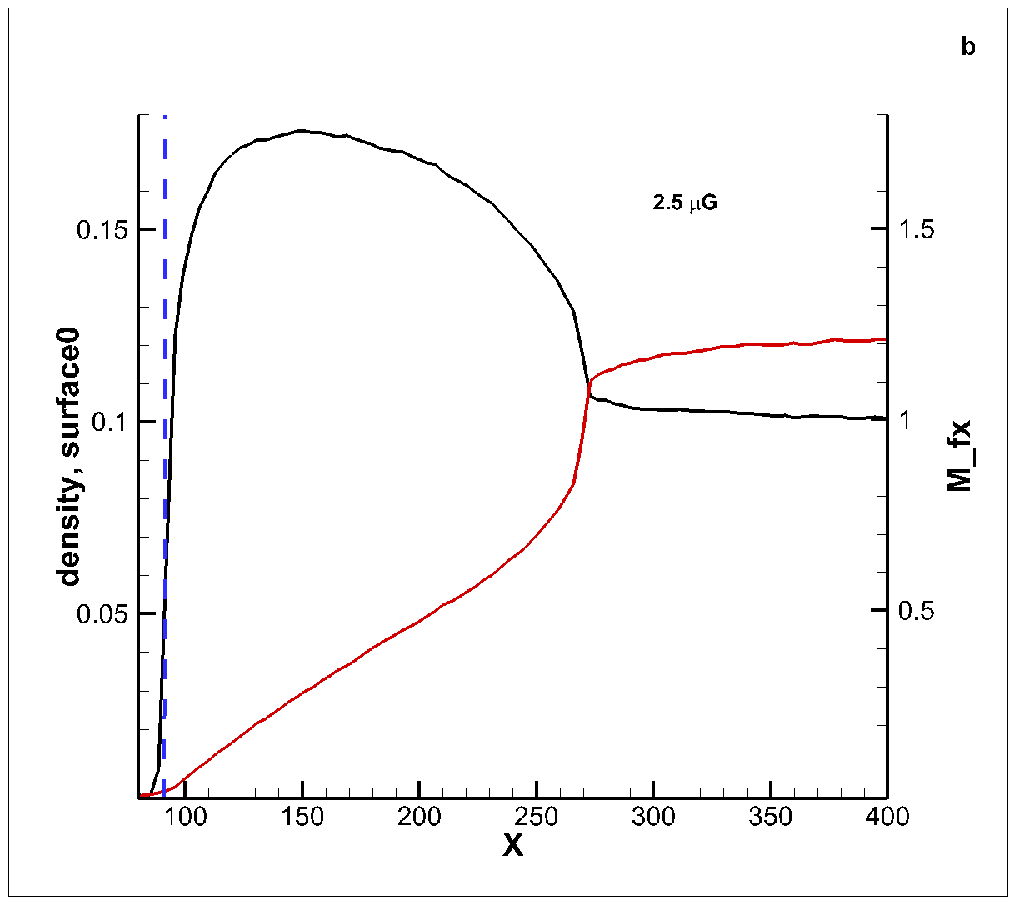}\\
\includegraphics[width=0.45\textwidth]{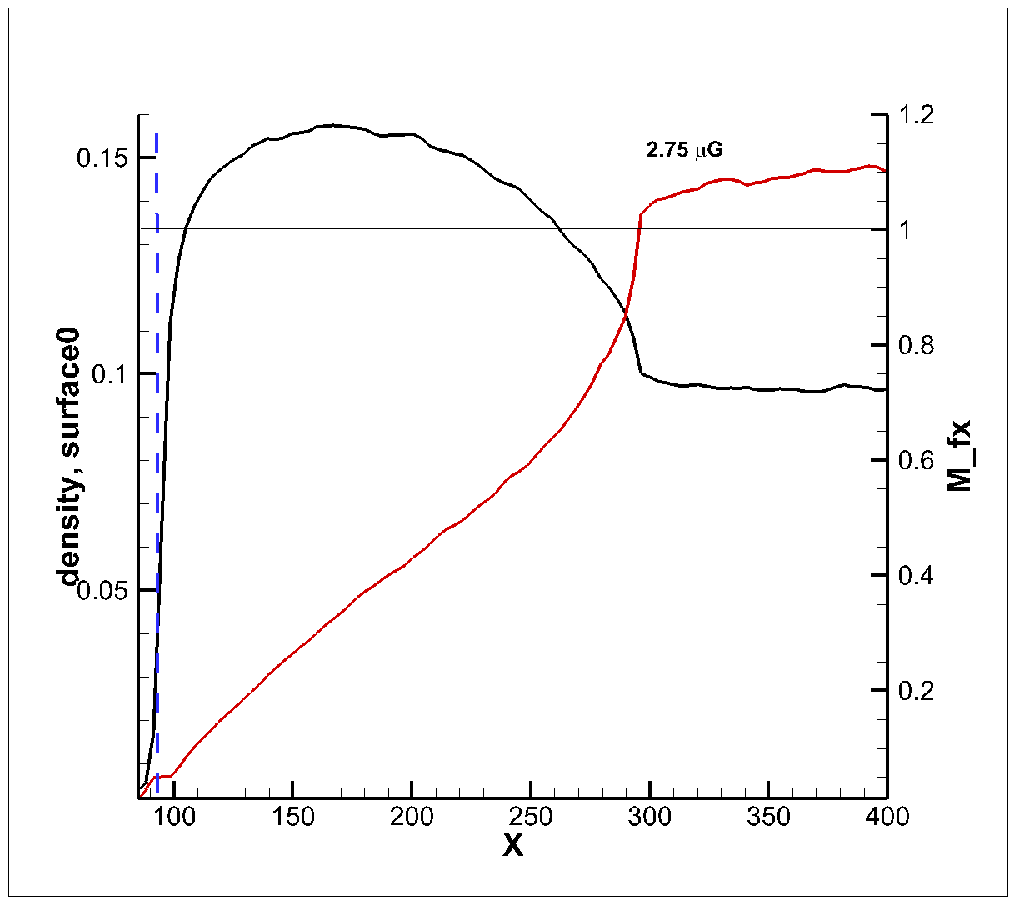}\hspace{5mm}
\includegraphics[width=0.45\textwidth]{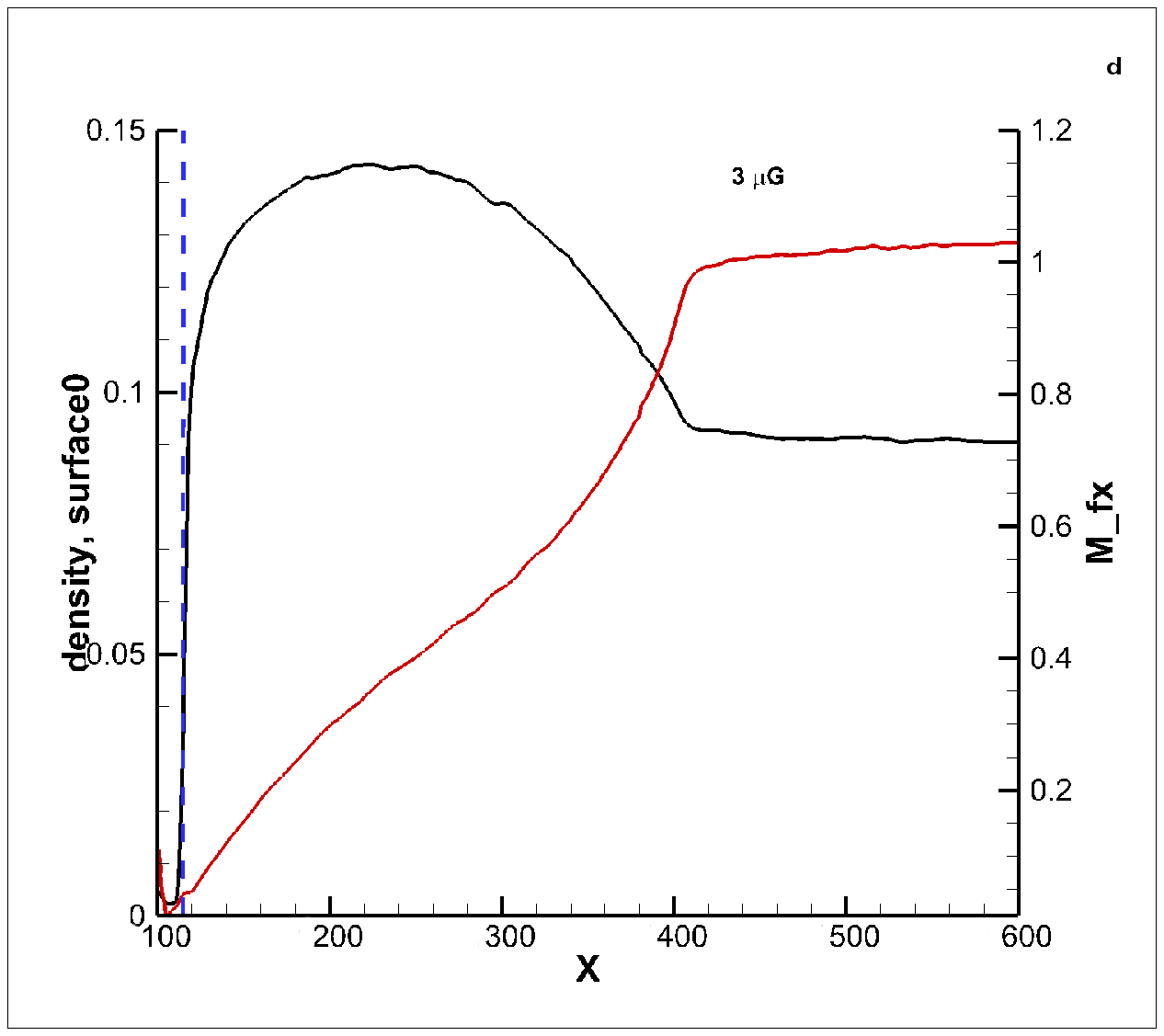}\\
\includegraphics[width=0.45\textwidth]{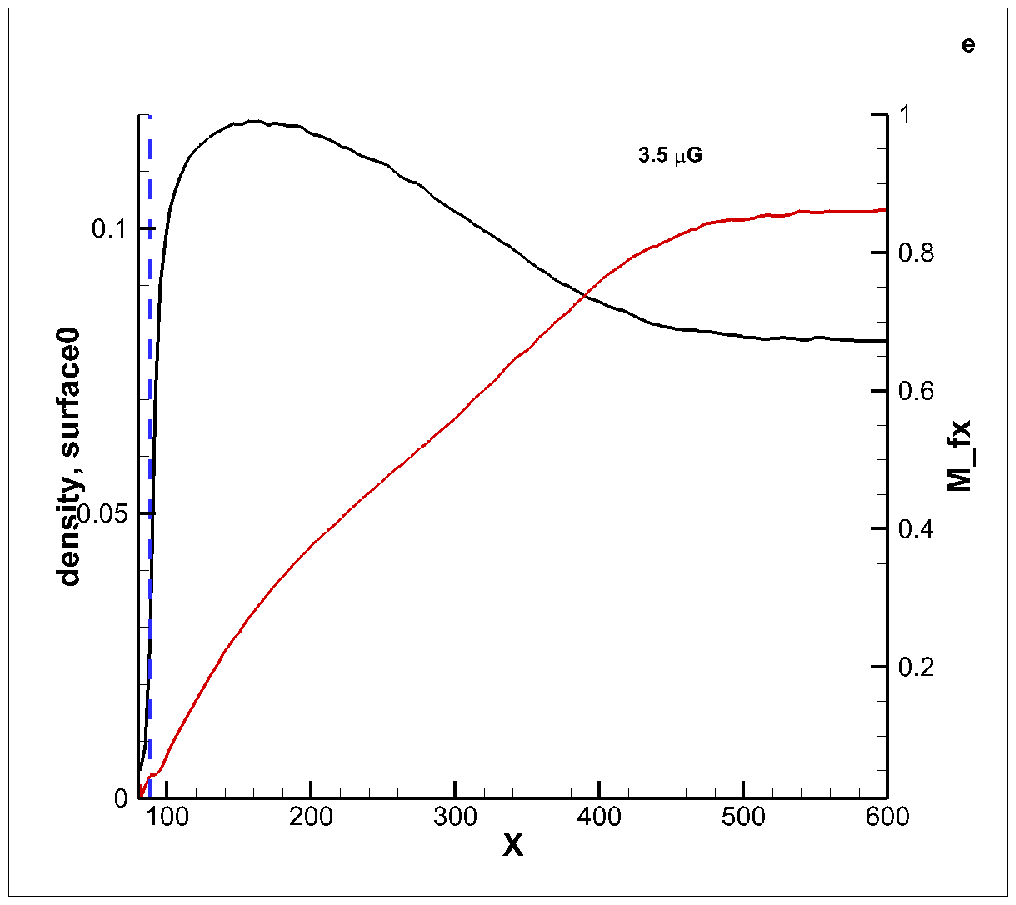}\hspace{5mm}
\includegraphics[width=0.45\textwidth]{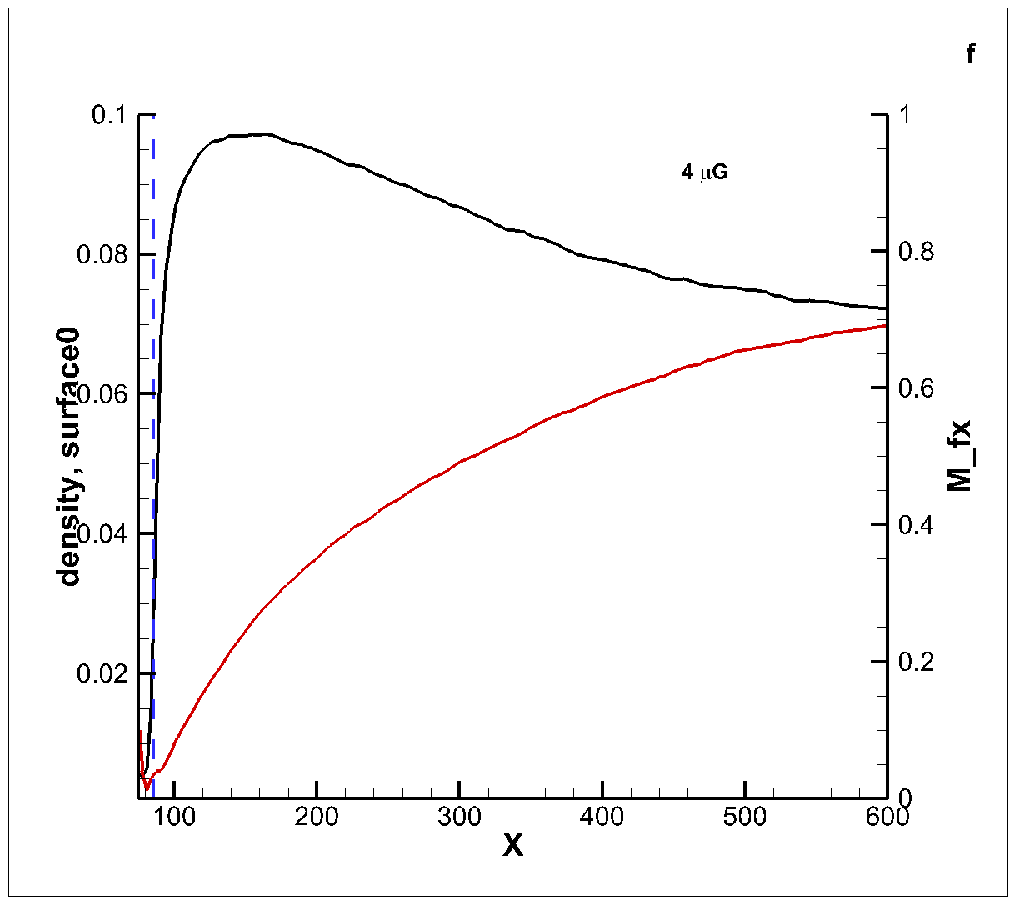}
\caption{Distributions of the plasma density (black lines) and fast magnetosonic Mach number (red lines)
along the $x$-axis in the simulations from Table~1. The vertical blue dashed lines shows the HP position.}
\label{fig2}
\end{figure*}

\begin{figure*}[p]
\centering
\includegraphics[width=0.45\textwidth]{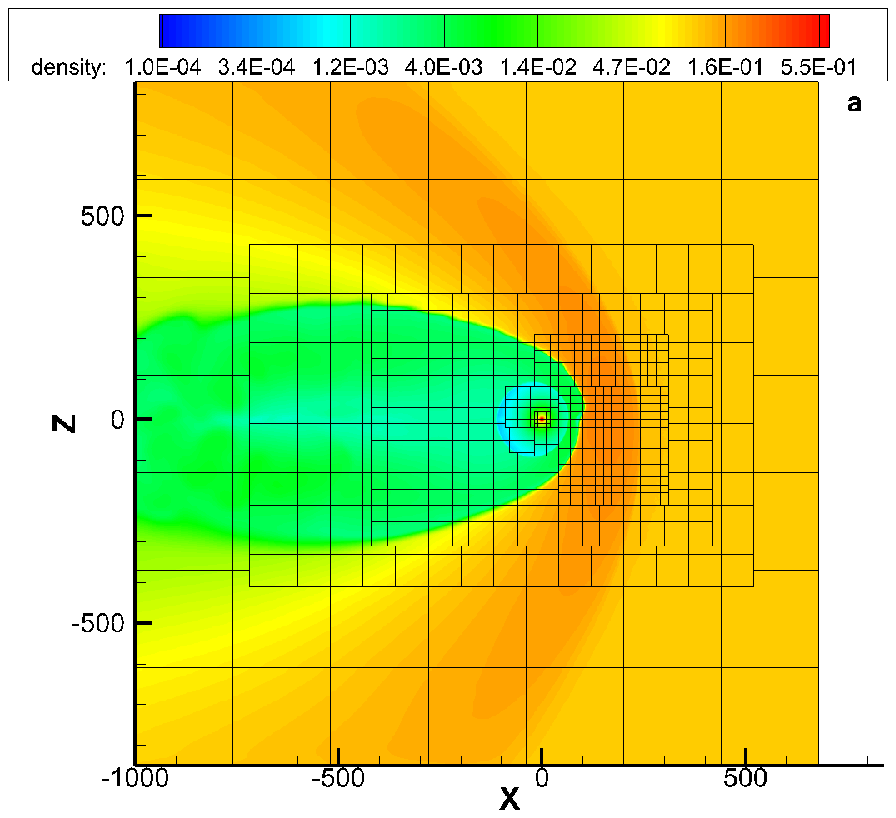}\hspace{5mm}
\includegraphics[width=0.45\textwidth]{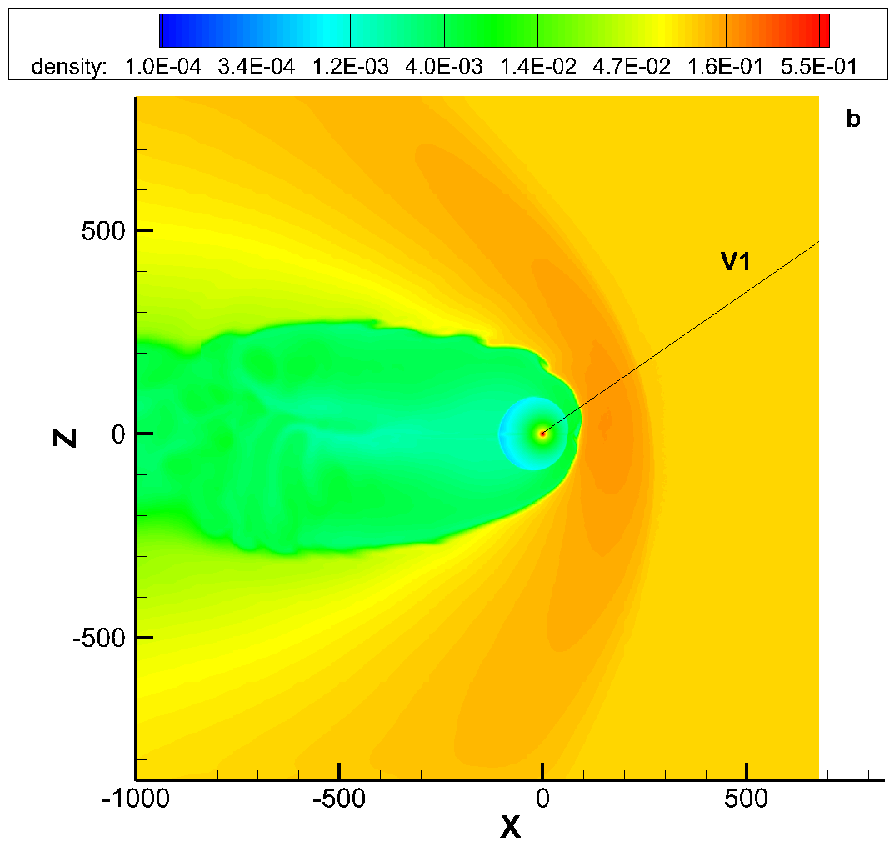}\\
\includegraphics[width=0.45\textwidth]{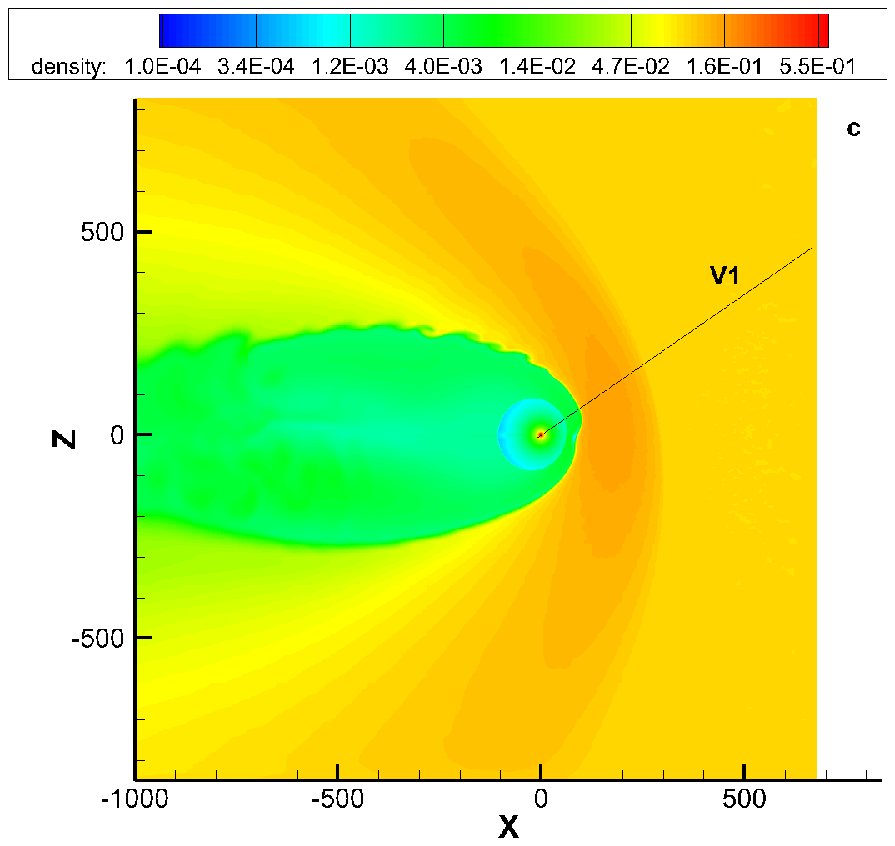}\hspace{5mm}
\includegraphics[width=0.45\textwidth]{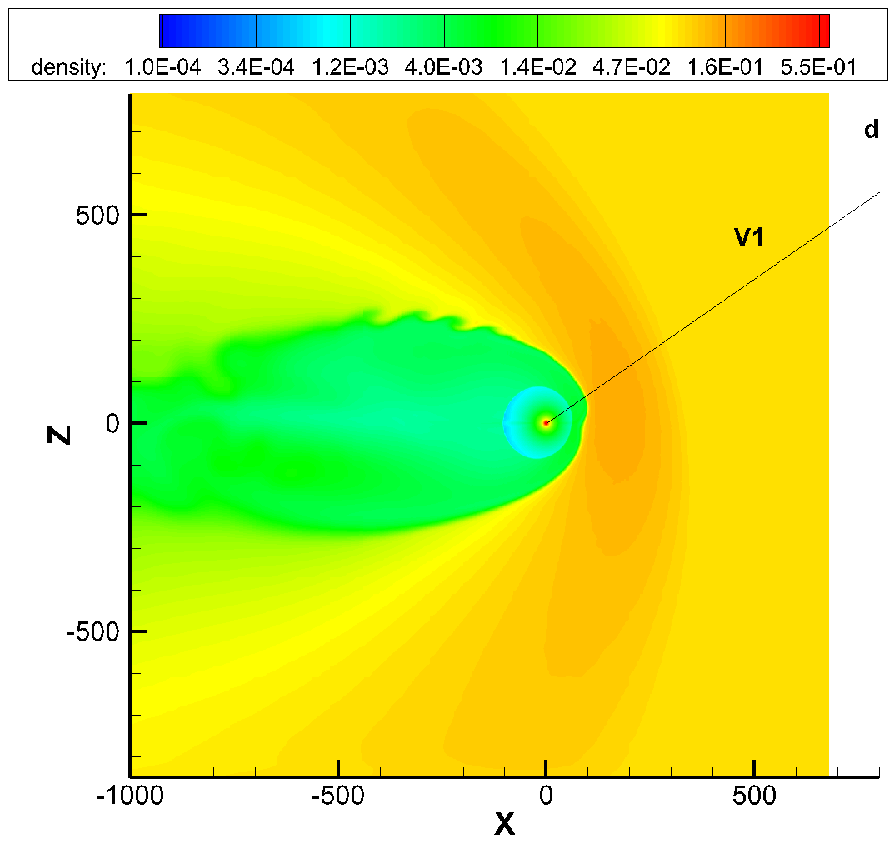}\\
\includegraphics[width=0.45\textwidth]{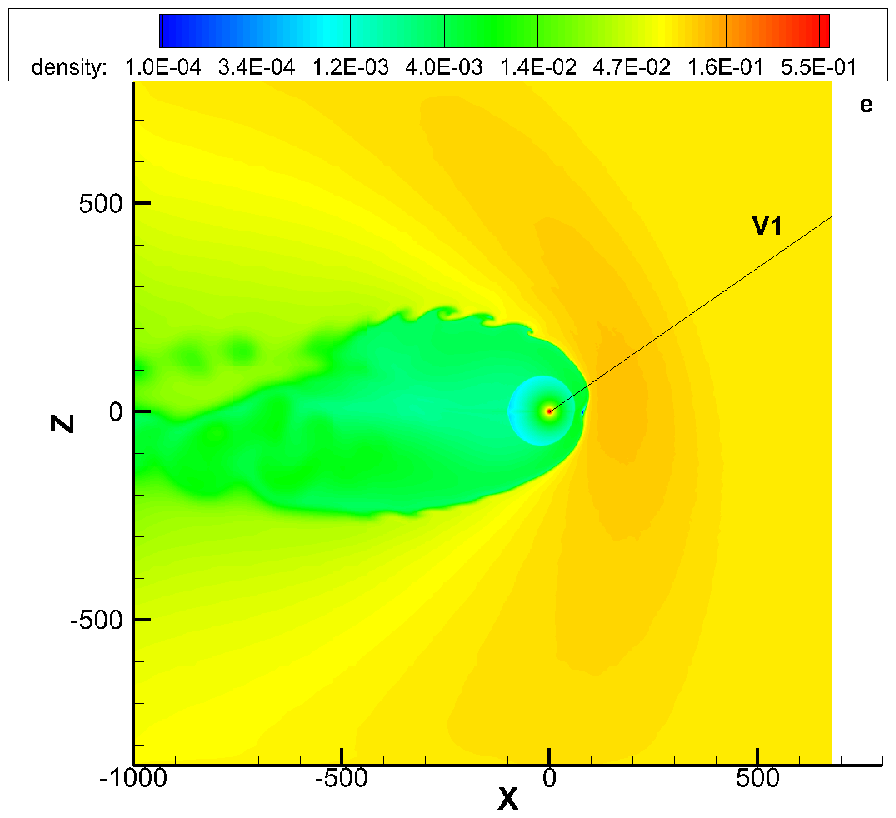}\hspace{5mm}
\includegraphics[width=0.45\textwidth]{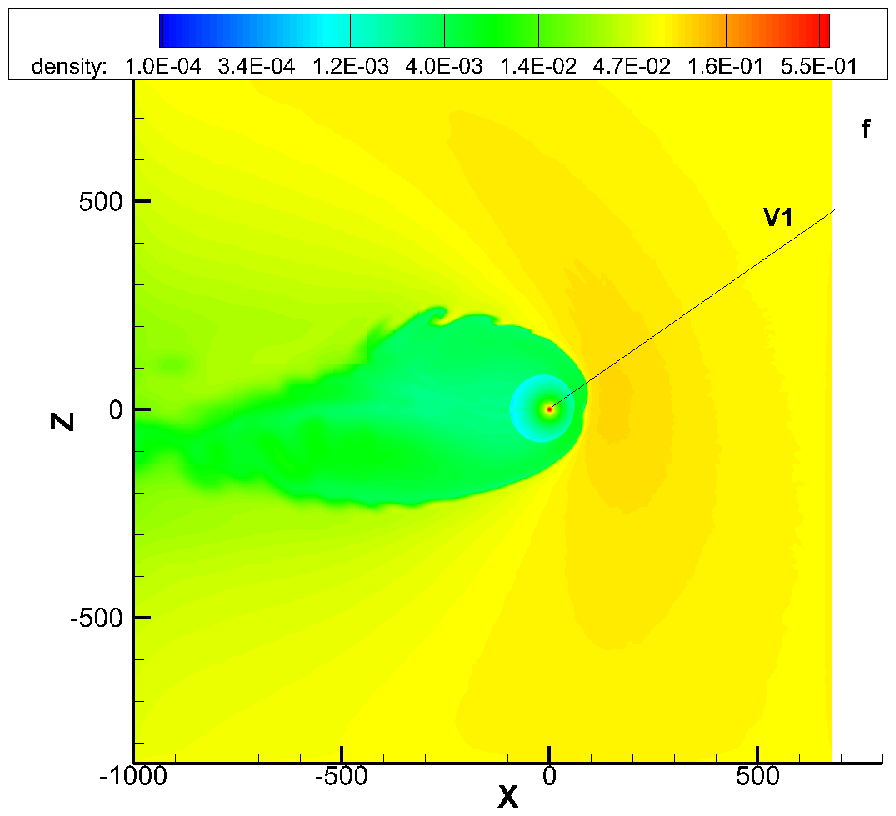}
\caption{Distributions of the plasma density in the meridional plane for the simulations from Table~1.}
\label{fig3}
\end{figure*}

Figure~\ref{fig2} shows the distributions of the plasma density and fast magnetosonic Mach number $M_\f= v/a_\f$ along the $x$-axis
behind the heliopause. Figure~\ref{fig3} shows the plasma density distributions for the same set of parameters in the meridional plane. The LISM parameters are taken from our previous simulation in \citet{Erik16}, and are summarized in Table~1. For convenience, the $\mathbf{V}_\infty$ and $\mathbf{B}_\infty$ directions are also given in the J2000 ecliptic coordinates (Table~2).
The SW properties are somewhat different from \citet{Erik16}, being closer to \textit{OMNI} data averaged over a substantial period of time 2120 days starting between 2010 DOY 1 and 2015 DOY 294) to obtain a spherically symmetric distribution at 1~AU: the plasma density is 5.924~cm$^{-3}$, velocity 409.8~km s$^{-1}$, temperature 82,336~K, and radial HMF component $39\ \mu$G.

It is clear from this figure that the LISM properties are substantially modified by charge exchange, the changes being stronger
behind the fast magnetosonic shocks seen in Figs.~\ref{fig2}a--\ref{fig2}b. This is not surprising since (1) the density of secondary neutral atoms affecting the LISM ions decreases with heliocentric distance and (2) the LISM plasma density increases as it approaches the heliospheric boundary layer (HBL) on the interstellar side of the HP.
It can be seen from Fig.~\ref{fig2}c that the shocked transition essentially disappears already at $B_\infty=2.75\ \mu$G: $M_\f\approx 1.02$ ahead of it. For $B_\infty =3\ \mu$G, $M_\f$ becomes smaller than 1 smoothly.
Further increase in $B_\infty$ makes the density variation from the unperturbed LISM to the HP weaker, but wider.
This effect is well-known \citep[see, e.g.][]{Izmod05,Pozaog06,Borov08,Zank10,Jacob15}. However, the solutions presented here are for the first time obtained using adaptive mesh refinement (AMR) in the OHS region (see the patch edges in Fig.~\ref{fig2}a), which made it possible to identify shocks inside the OHS plasma and the HBL near the HP. In some of presented simulations, such shocks are situated inside regions of substantial variation of the LISM properties.
From this viewpoint, the OHS itself can be called a ``bow wave,'' which forms in front of the HP due to the SW--LISM collision. This bow wave can either have a shock inside it or not, depending on the full set of SW and LISM parameters. The presence of a shocked transition inside the bow wave clearly is not determined by the condition $M_\f>1$ in the unperturbed LISM plasma. To illustrate the spatial extent of the bow wave, in  Figure~\ref{fig3}, we show the plasma density distributions in the meridional plane for all simulations described in Table~1.

Our simulations also make it possible to understand the nature of the ``boundary layer.'' This layer reveals itself as a plasma density decrease in front of the HP. In previous simulations, numerical smearing of the HP made it difficult to distinguish the density increase from the solar side to the LISM across the HP itself and the density increase following it. Boundary layers are formed also upstream of the
Earth's magnetopause \citep{Zwan}, where they are called the plasma depletion layers (PDLs). Following \citet{Lees} and \citet{Alksne}, it was shown that a PDL on the surface of the magnetosphere creates magnetic stress that affects the plasma flow. The width of such depletion layer was estimated to be about 700--1300 km for the Earth magnetosphere at 10~$R_\mathrm{E}$ for the SW Alfv\'en number $M_\A=8$ and rapidly decreasing as $M_\A$ increases. When simplistically scaled to the size of the outer heliosphere, a depletion layer at the HP would be 1\%--2\% of the heliocentric distance of the latter, which gives us about 1.25--2.5~AU. The ions of the terrestrial magnetosheath are typically observed to have bi-Maxwellian velocity distributions with $T_\perp /T_\parallel >1$, where the superscripts $\perp$ and $\parallel$ denote directions perpendicular and parallel to the background magnetic field. \citet{Anderson} showed that the temperature anisotropy may be important for the PDL formation because it can launch an electromagnetic ion cyclotron instability, which makes scattered ions propagate along the magnetic field and leave the draping region. In the more recent analysis of \citet{Fuselier}, the PDL width is in the range 1--5~AU, which emphasizes the necessity to incorporate microphysical processes of the magnetic field draping/depletion layer formation into the PDL analysis \citep[see also][]{Gary,Denton}. It is interesting to notice, however, in this connection that a HBL exists in on the LISM side of the HP in simulations without magnetic field \citep{Bama93}, which means that charge exchange affects them somehow. To separate the density decrease  from the LISM side to the SW side of the HP, it is necessary either to fit the HP surface ensuring the satisfaction of the boundary conditions suitable for tangential discontinuities in MHD, or use AMR. \citet{Izmod15} show that such boundary layer exists also on the SW side of the HP. It is interesting to note in this connection that \citet{Beru} argue that the density jump across the HP may disappear at the LISM stagnation point, the density variation being smooth and occurring mostly in the boundary layers.

\begin{figure}[t]
\centering
\includegraphics[width=\columnwidth]{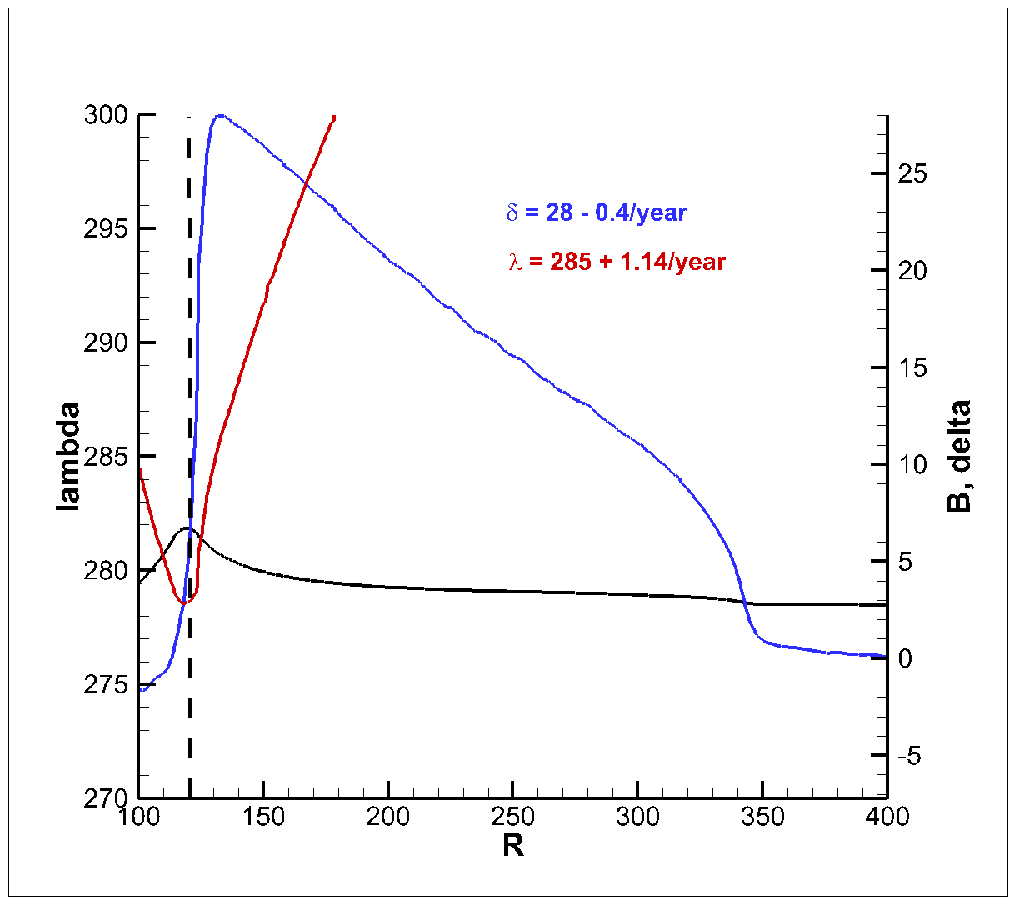}
\caption{Distributions of the magnetic field vector magnitude $B$ (black line) and its elevation (blue line) and azimuthal (red line) angles,
$\delta$ and $\lambda$, along the Voyager~1 trajectory.}
\label{fig4}
\end{figure}

It is seen from Figs.~\ref{fig2} that the LISM plasma density reaches its maximum at 50--100 AU from the HP surface, which is of the order of 1--2 charge exchange free paths in the OHS. The maximum values are: (1) 0.2; (2) 0.175; (3) 0.155; (4) 0.14;
(5) 0.115; and (6) 0.095 cm$^{-3}$, respectively. Further on, the plasma density is only decreasing in the sunward direction, and the maximum gradient is at the HP surface. Since we determine the HP position exactly, by solving a level-set equation for the boundary between the SW and the LISM
\citep{Borov11}, we can look closer at the magnetic field  and  density variations in the HBL. In Fig.~\ref{fig4}, we show the magnetic field vector magnitude, $B$,
and its elevation and azimuthal angles, $\delta$ and $\lambda$, along the \textit{V1} trajectory in the simulation with $B_\infty=2.75\ \mu$G from Table~1. In agreement with observations, both $B$ and $\lambda$ are continuous across the heliopause. The grid resolution near the HP is 1.2~au cubed. The HP position is shown with the vertical dashed line. The numerical smearing of quantities is about $\pm 3$~au near the HP.
Magnetic field strength slightly increases inside the HBL. On the other hand, the angle $\delta$ increases to about $28^\circ$. The observed values are: $\delta=22^\circ \pm 3^\circ$ and $\lambda=291^\circ \pm 3^\circ$ \citep{B1}. In addition, the simulation shows a consistent undraping of the ISMF as \textit{V1} propagates deeper into the LISM. The gradient in $\delta$ and $\lambda$ are consistent with observations reported by \citet{Burlaga14}. However, these gradients  become smaller if averaged from the crossing time to 2016 \citep{Burlaga16}.  This leads us to a conclusion that time-dependent phenomena, such as described in  \citet{Fermo} are likely to affect the undraping.

\begin{figure*}[t]
\centering
\includegraphics[width=0.48\textwidth]{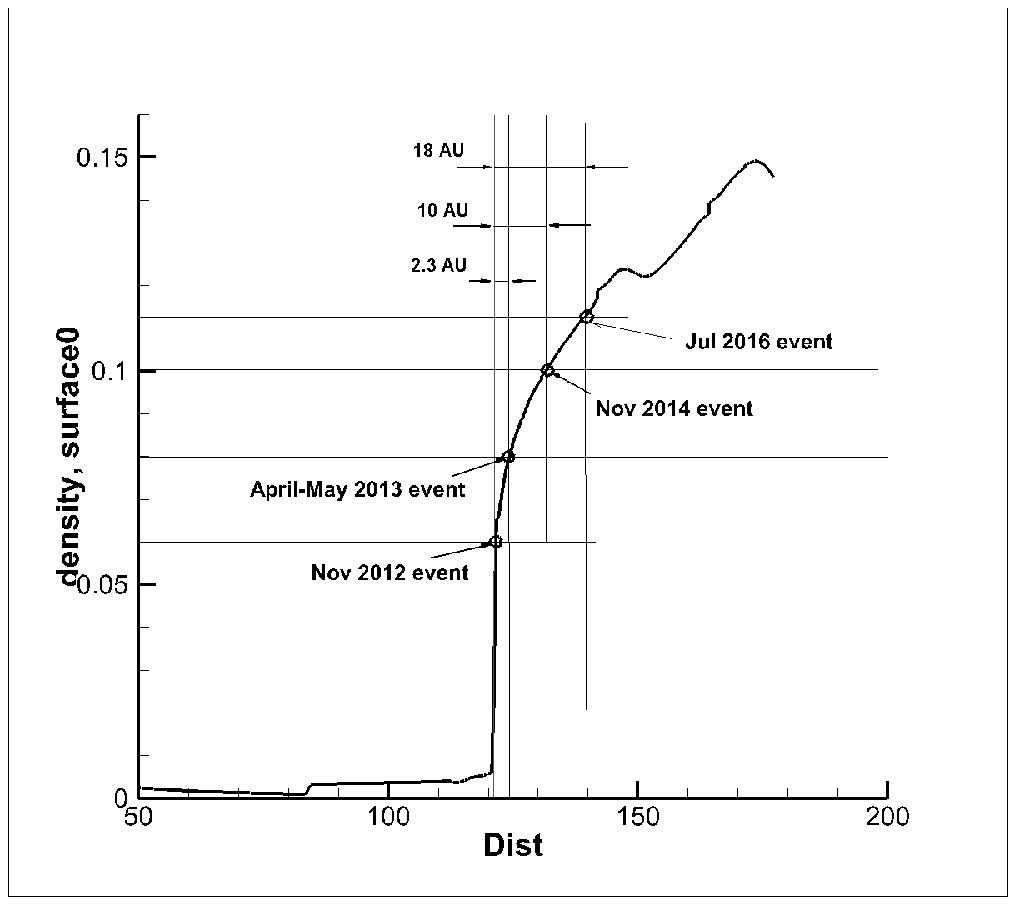}
\includegraphics[width=0.48\textwidth]{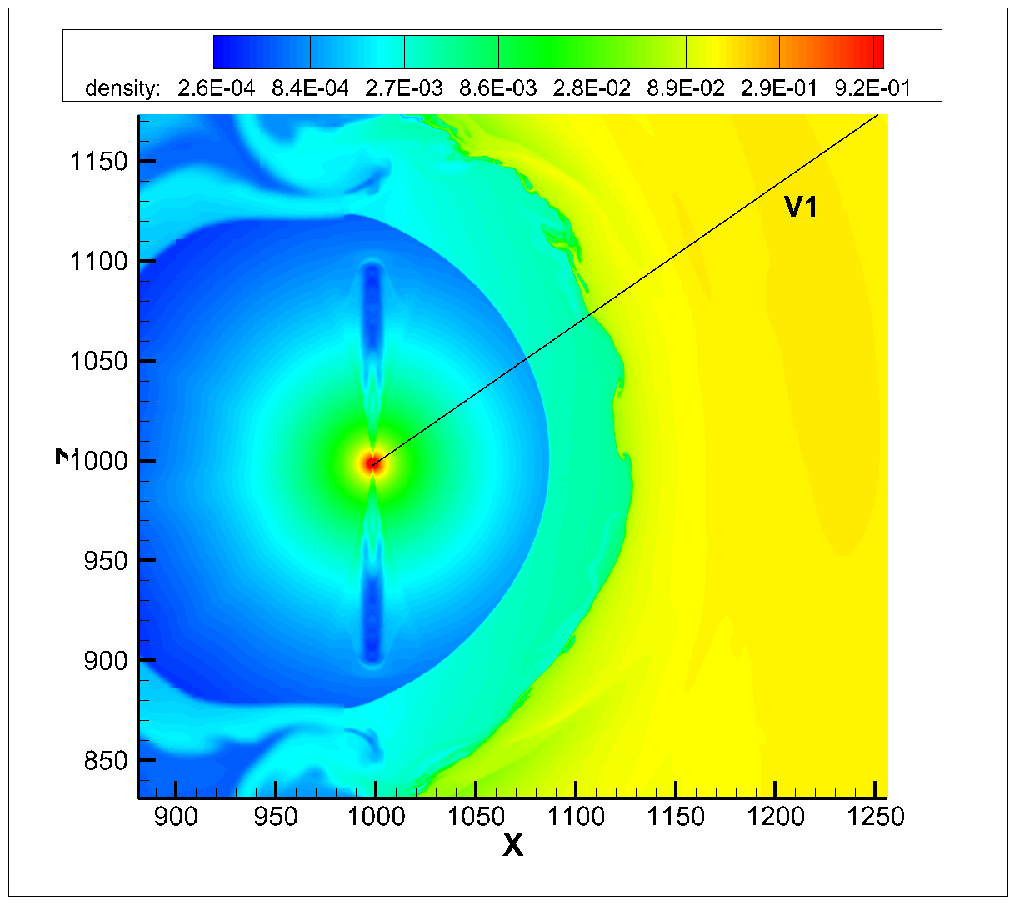}
\caption{The distribution of plasma density \emph{(left panel)} along the \textit{V1} trajectory and its comparison with the plasma waves events detected by the spacecraft beyond the heliopause, and (\emph{right panel}) in the meridional plane.}
\label{fig5}
\end{figure*}

\begin{figure*}[t]
\centering
\includegraphics[width=0.8\textwidth]{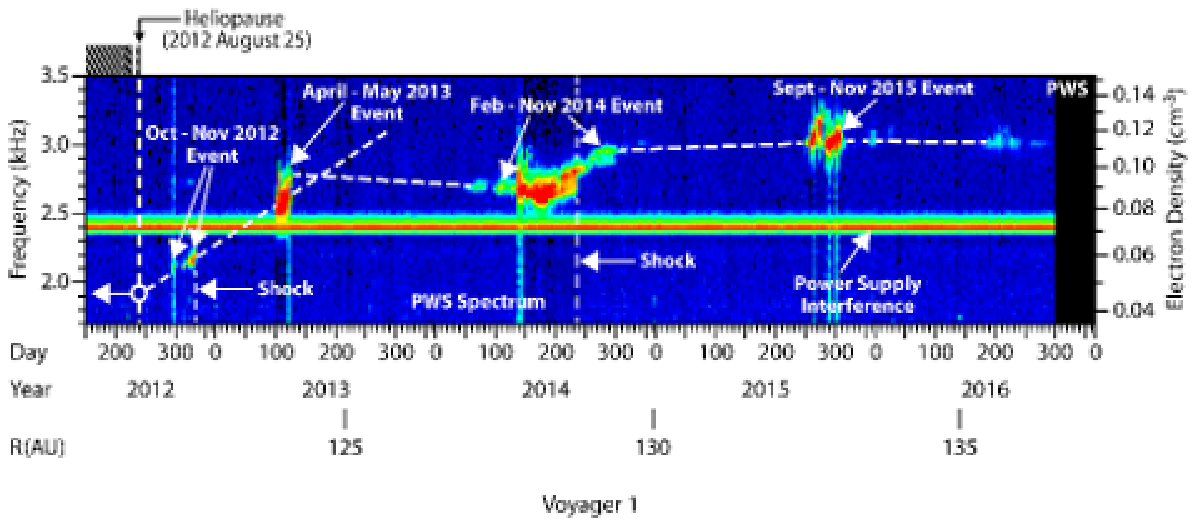}
\caption{\emph{Voyager}~1 measurements of plasma oscillation frequency and electron density derived from it.}
\label{Gurnett}
\end{figure*}

Figure~\ref{fig5} shows the plasma density distribution in the meridional plane (\textit{right panel}) and along the \textit{V1} trajectory
(\emph{left panel}) in a time-dependent simulation which particularly focuses on the heliopause resolution. The LISM parameters are taken from Table~1 ($B=3\ \mu$G), but a nominal (periodic with the 11-year period) solar cycle is taken into account, similarly to \citet{Pogo09a} and \citet{Borov14}. The length scale has been decreased by a factor of 1.1 on the left panel, to assure the HP position in the observed point.
The density increase with distance from the heliopause should result in the increase of the electron oscillation plasma
frequency, in accordance with \textit{V1} observations with the Plasma Wave Instrument (PWS) \citep{Gurnett15} (see also Fig.~\ref{Gurnett}). Initially, the density increased from 0.06~cm$^{-3}$ to 0.08~cm$^{-3}$ from from Nov 2012 to April-May 2013 (on the distance of $\sim 2$~AU traveled by \textit{V1}).) In the figure, this distance is somewhat greater ($\sim 2.3$~AU). The next wave emission event measured by \textit{V1} occurred in November 2014
and showed the density increase to about 0.09--0.11~cm$^{-3}$. The spacecraft traveled approximately 7~AU between these events. The simulation shows the distance of about 10~AU. Two more plasma oscillation events were measured by PWS: on Sep--Nov 2015, when the density measured on Day 298, 2015 was 0.115~cm$^{-3}$ at a distance of about 133 AU from the Sun, and on Day 201, 2016, when the density was determined to be
0.113~cm$^{-3}$ at a distance of 135.5~AU. We show the latter event in Fig.~\ref{fig5}, where it occurred $\sim 139$~AU.
In addition, the measured density change between the latest two events was almost negligible, which is not seen in our simulation.
These discrepancies should not be surprising because our model is not detailed enough to identify MHD shocks propagating through the OHS due to realistic time-dependent boundary conditions.  The figure does show, however, the perturbations propagating through the LISM due to the solar cycle. It can be seen that there are intervals as large as 5--6~AU without any density increase. The density gradient becomes smaller as \textit{V1} continues to traverse the LISM and it may take $\sim 9$~yrs until it reaches the maximum density value.


\section{Instabilities and magnetic reconnection near the heliopause}
As mentioned above, the HP instability may be responsible for the HP ``structure'' observed by \textit{V1} before it entered the LISM completely. The HP is also a likely venue for magnetic reconnection.
It is known from both theory and simulations  that the HP is unstable both at its nose and on the flanks \citep[see][and references therein]{Ruderman10,Pogo14,Pogo17,Avinash}. Charge exchange between ions and neutral atoms play an important role here through the action of the source terms in the momentum and energy equations. Near the stagnation point, where the shear between the SW and LISM flow is small, charge exchange results in a sort of Rayleigh--Taylor (RT) instability \citep{Liewer,Zank99,Florin05,Borov08b}, which is known to take place when a heavier fluid lies upon a lighter one. Farther from the stagnation point, the Kelvin--Helmholtz (KH) and other instabilities may develop \citep{Ruderman15}.
The instabilities of the HP nose are seen only at high numerical resolution in this region ($\sim 0.1$~AU in the simulation shown below in Fig.~\ref{fig6}), which is too expensive computationally for an MHD-kinetic model. This is why, we use a multi-fluid model here with 4 neutral fluids involved.
It is interesting to see an agreement between simulations of \citet{Borov08b} and analytic analysis of \citet{Ruderman15} regarding the HP instability on its flanks. Both analyses demonstrate that charge exchange is primarily responsible for the flank destabilization, whereas it is further influenced by the shear flow in the HP vicinity. As shown by \citet{Borov14}, the HMF can partially stabilize the HP in the nose. However, previous solar-cycle simulations \citep{Pogo09a,Pogo13b} demonstrate that the HMF indeed becomes small near the HP at certain stages of the solar cycle
(see also Fig.~\ref{fig6}). This can be understood if we realize that the SW region swept by the HCS always embraces the solar equatorial plane before it reaches the TS. The SW streamlines that start in this region are directed towards a vicinity of the stagnation point on the inner side of the HP. As a result, the SW streamlines that carry magnetic field depressed by the processes occurring in the HCS-covered region of the IHS should spread over the HP surface \citep{Pogo14}. Although it is shown in \citet{Borov14} that solar cycle itself is not required for the RT-instability to develop, time dependencies in the SW do affect the actual evolution of such instabilities.
\begin{figure*}[p]
\centering
\includegraphics[width=0.45\textwidth]{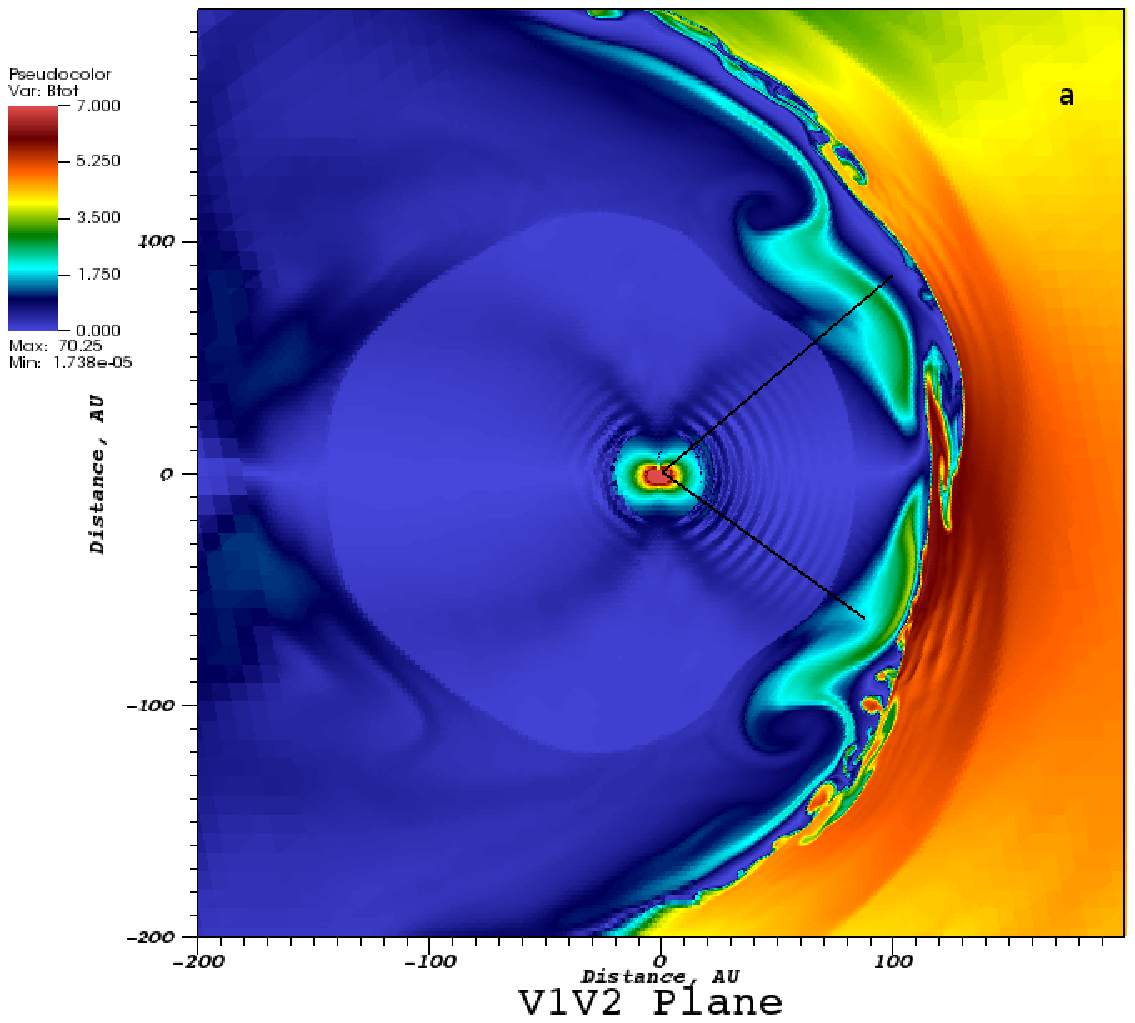}\hspace{5mm}
\includegraphics[width=0.45\textwidth]{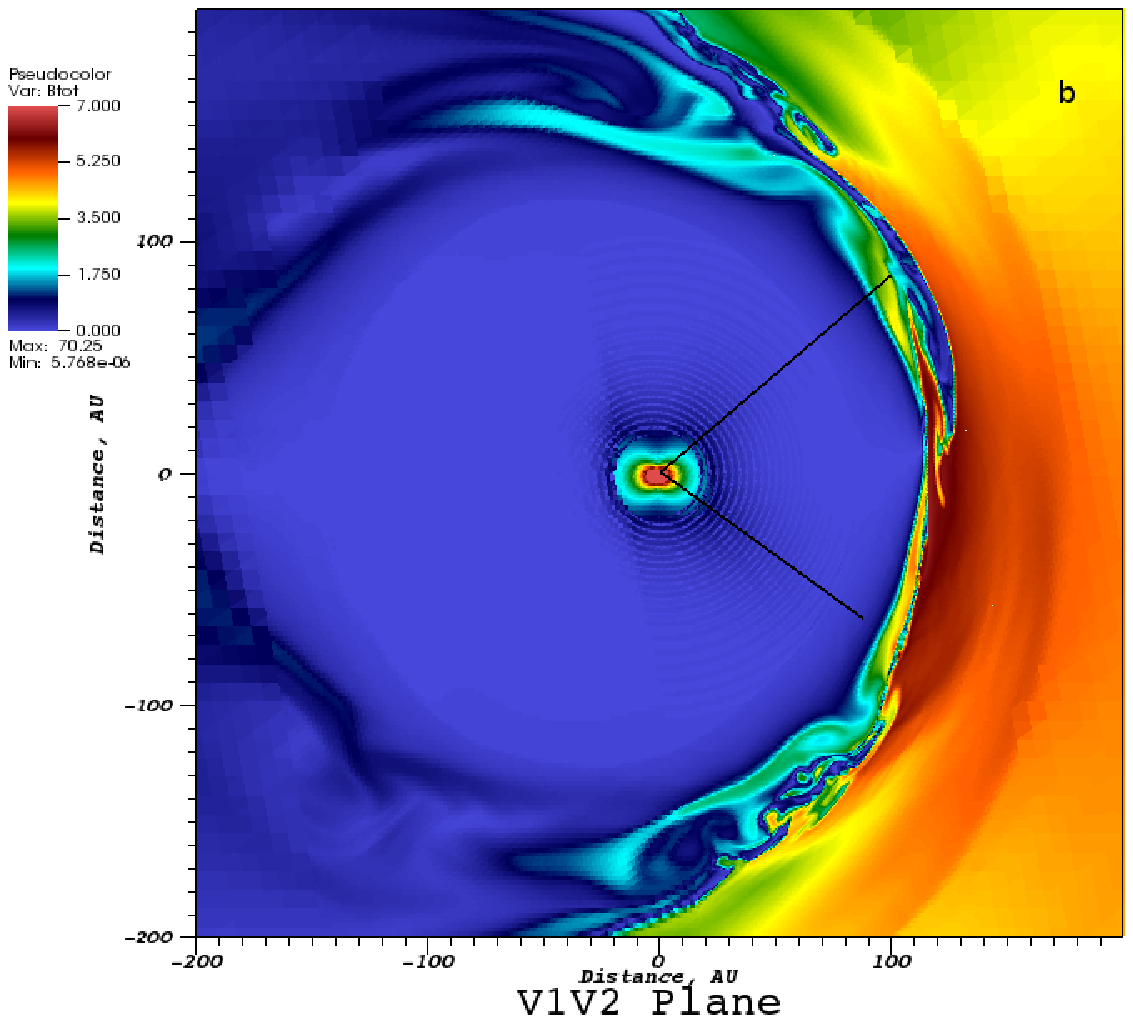}\\
\includegraphics[width=0.45\textwidth]{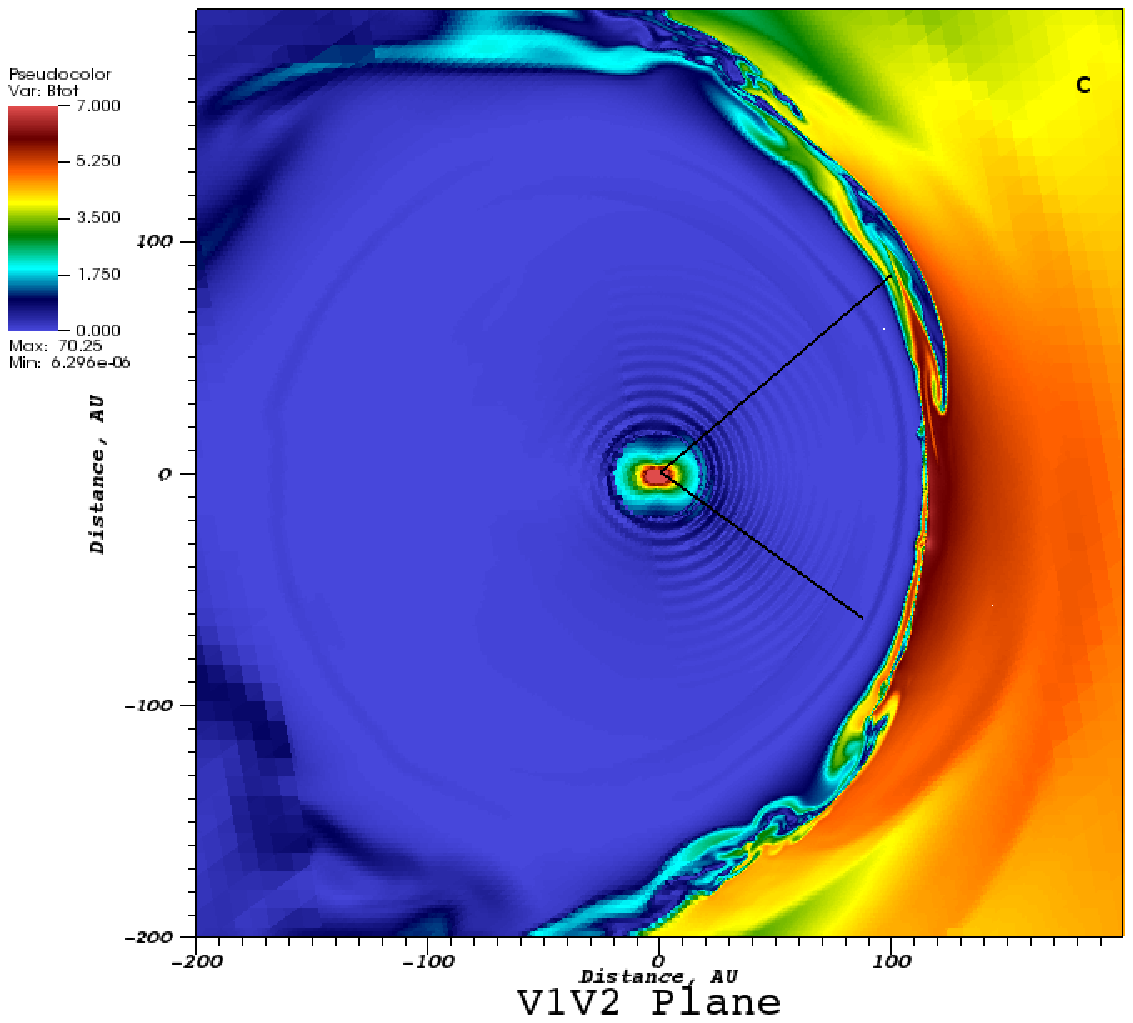}\hspace{5mm}
\includegraphics[width=0.45\textwidth]{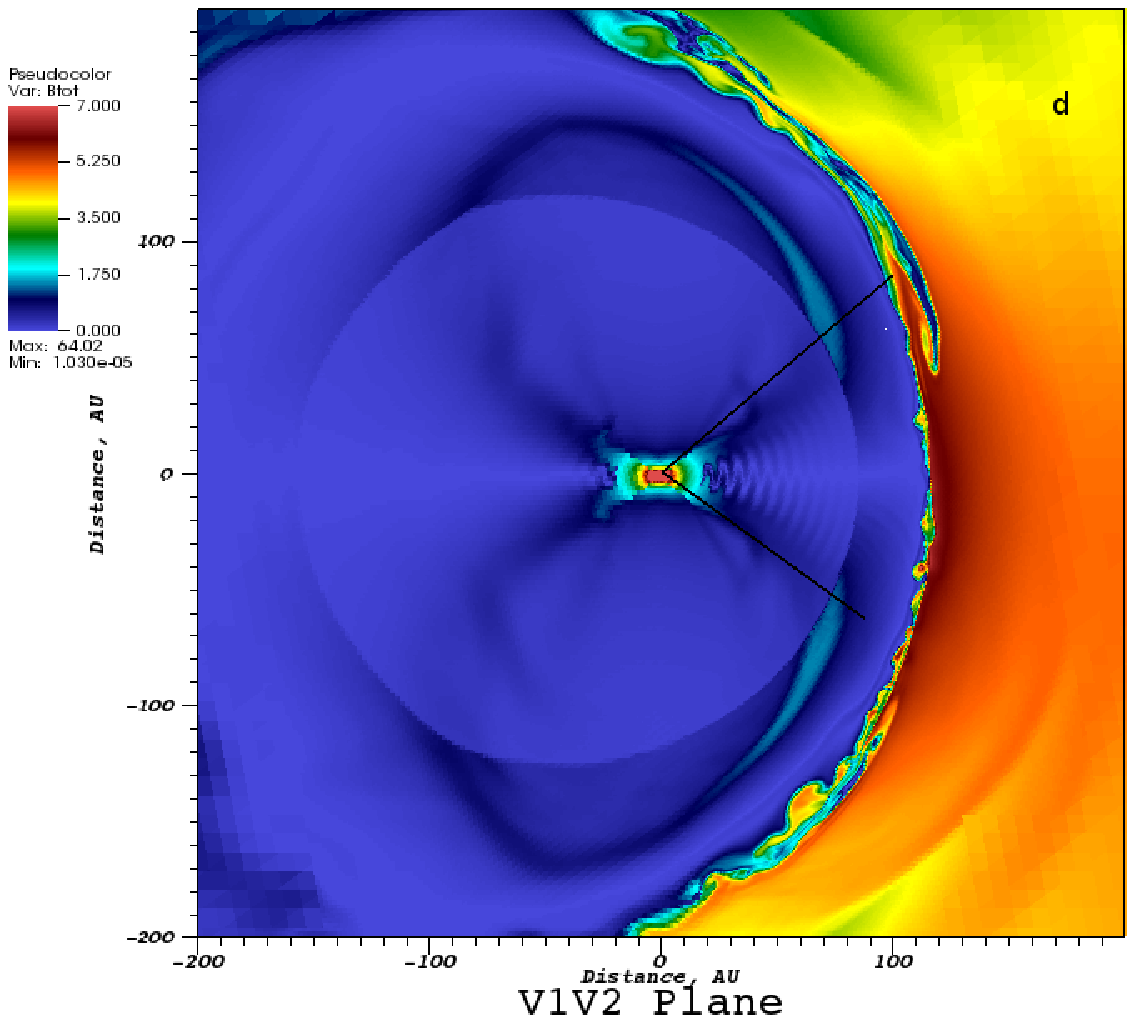}\\
\includegraphics[width=0.45\textwidth]{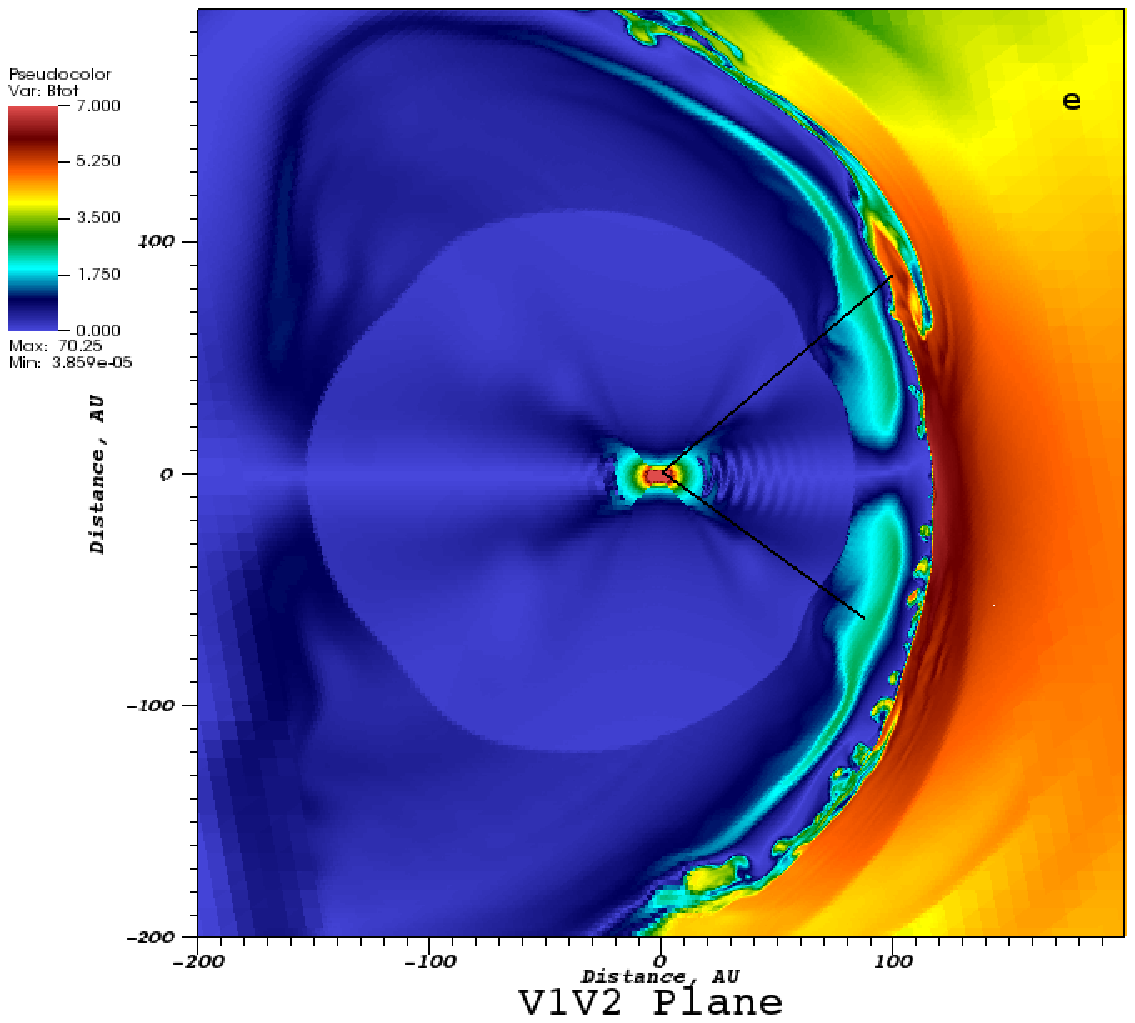}\hspace{5mm}
\includegraphics[width=0.45\textwidth]{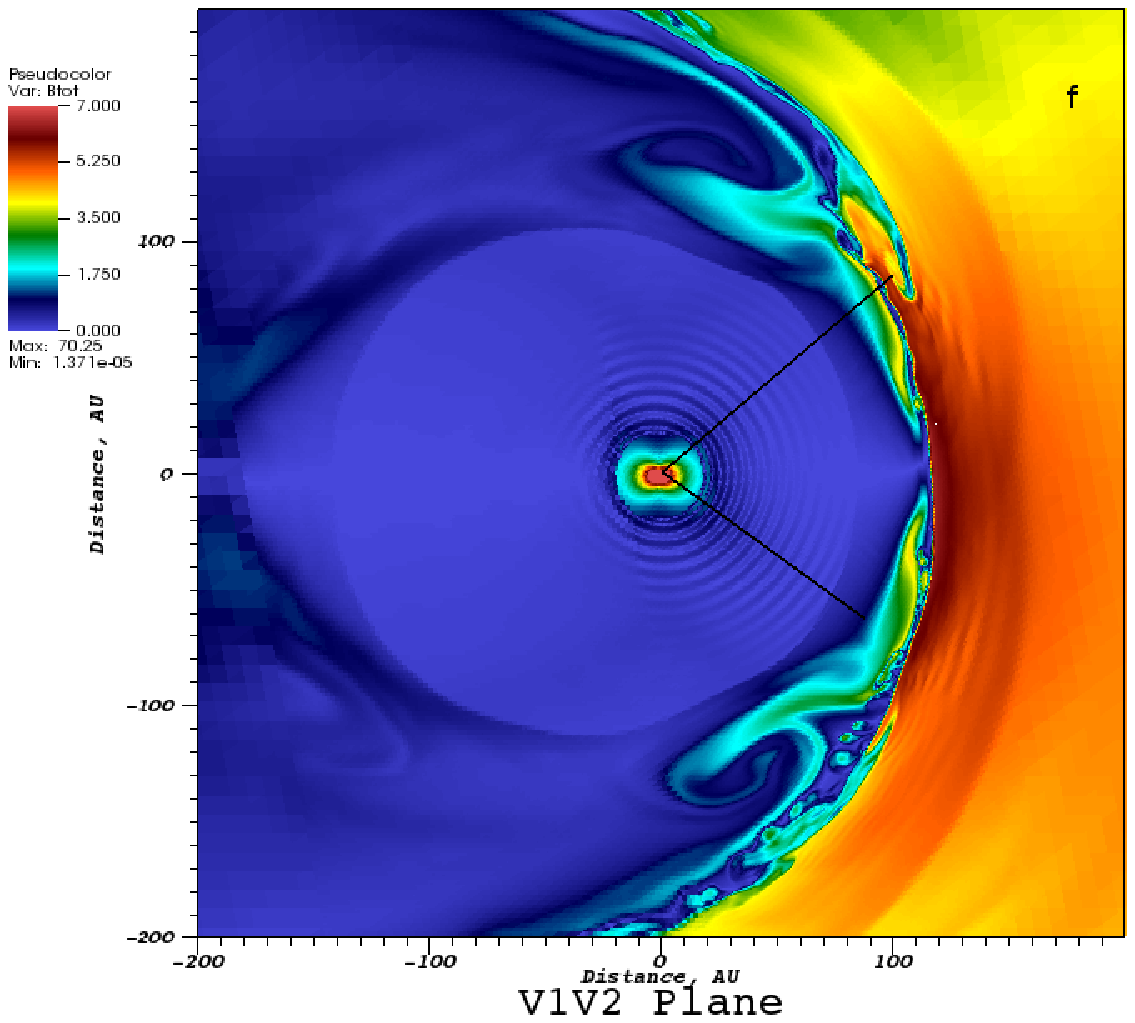}
\caption{Magnetic field magnitude behavior in the plane formed by the \textit{V1} and \textit{V2} trajectories
in the simulation that takes into account solar cycle effects. Time is increasing from the left to the right and from the top to the bottom.}
\label{fig6}
\end{figure*}
\begin{figure*}[p]
\centering
\includegraphics[width=0.45\textwidth]{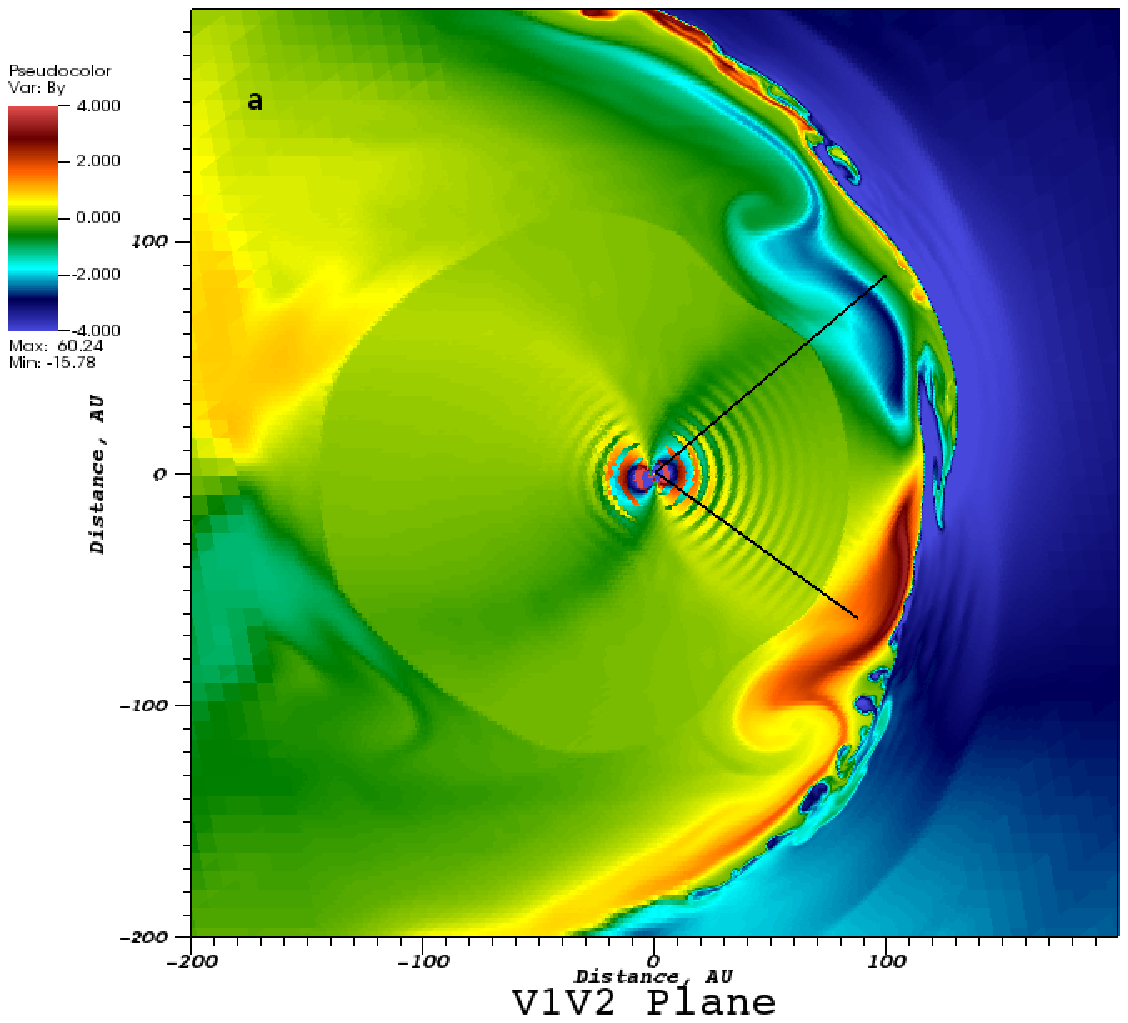}\hspace{5mm}
\includegraphics[width=0.45\textwidth]{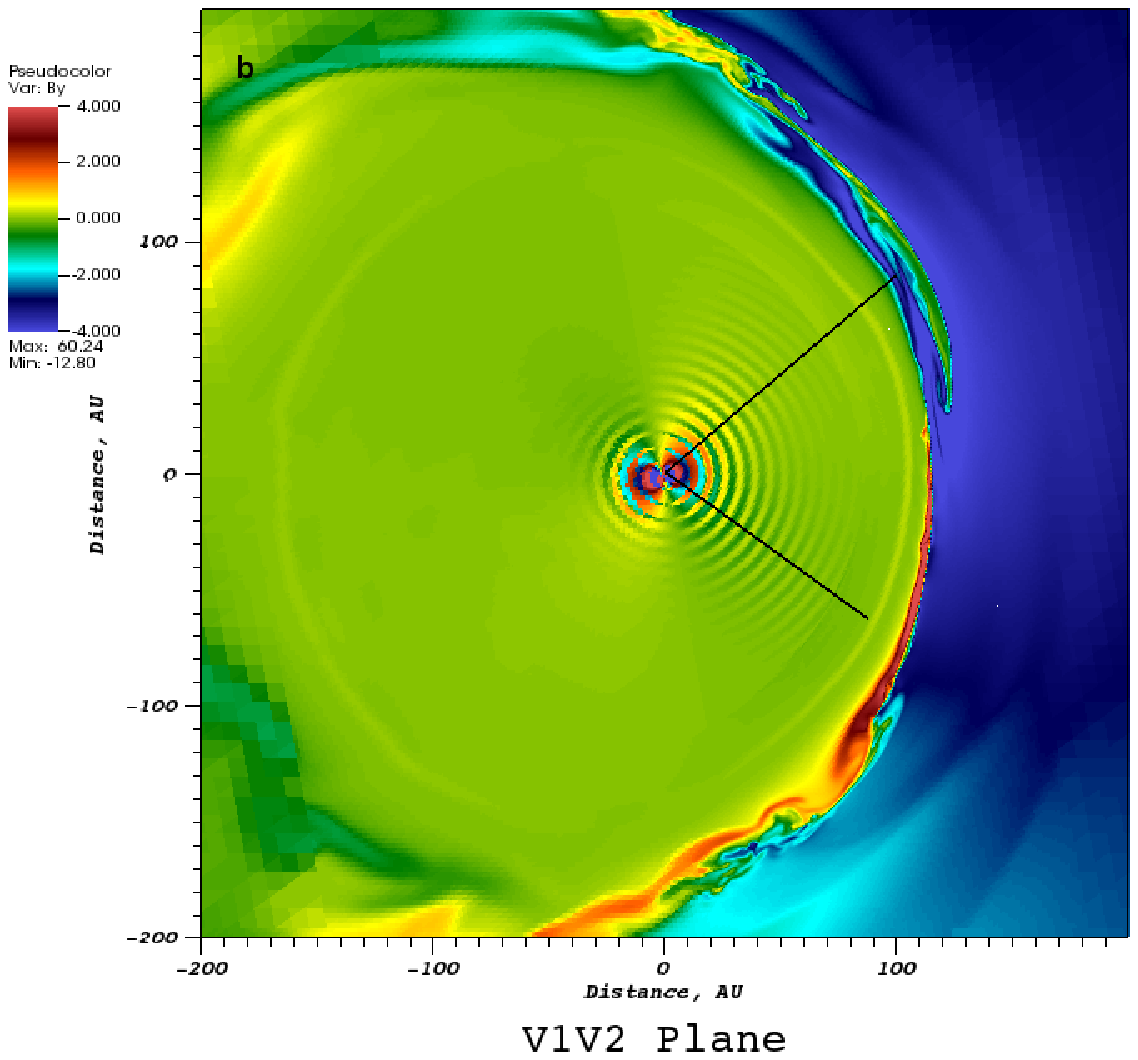}\\
\includegraphics[width=0.45\textwidth]{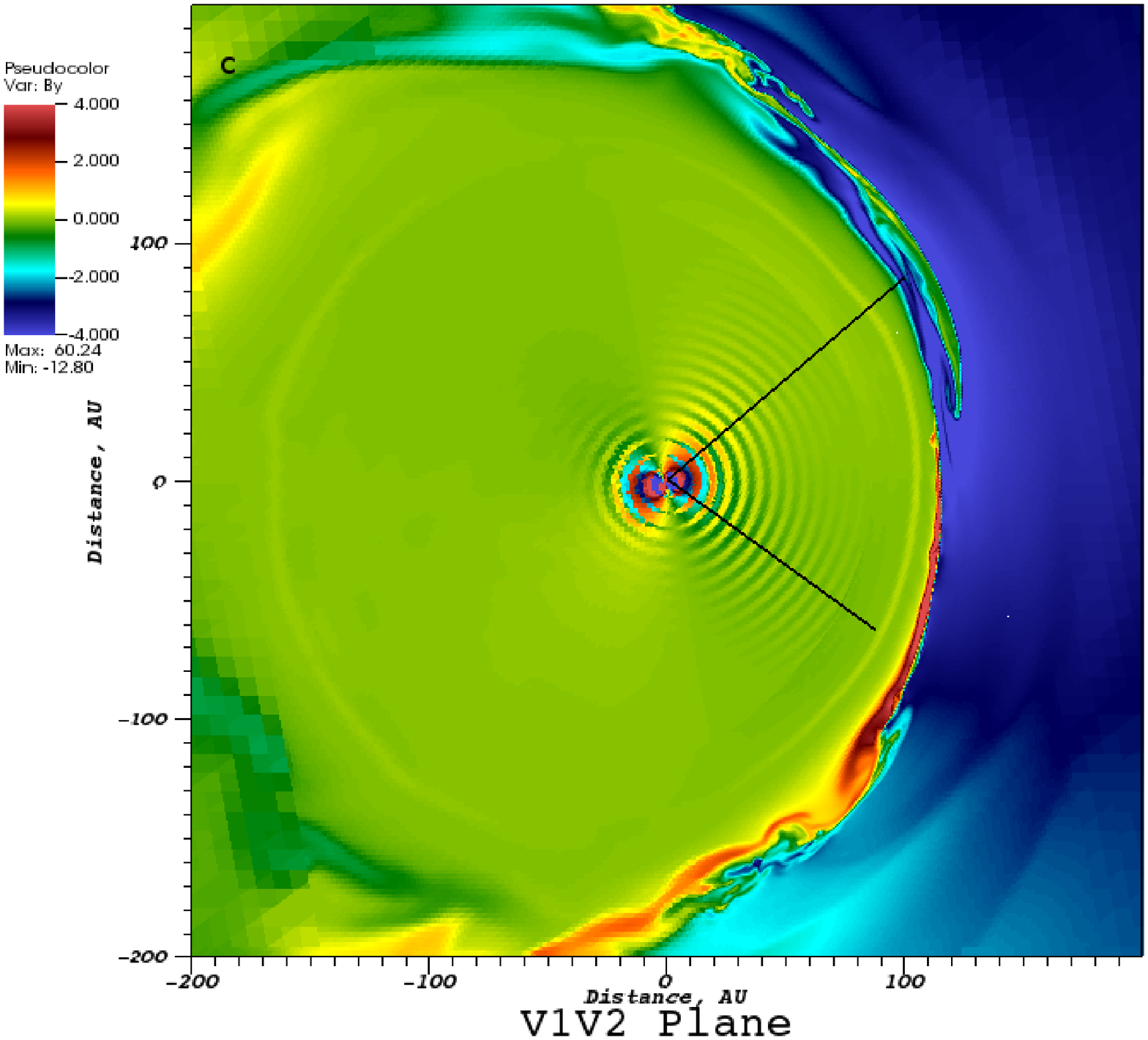}\hspace{5mm}
\includegraphics[width=0.45\textwidth]{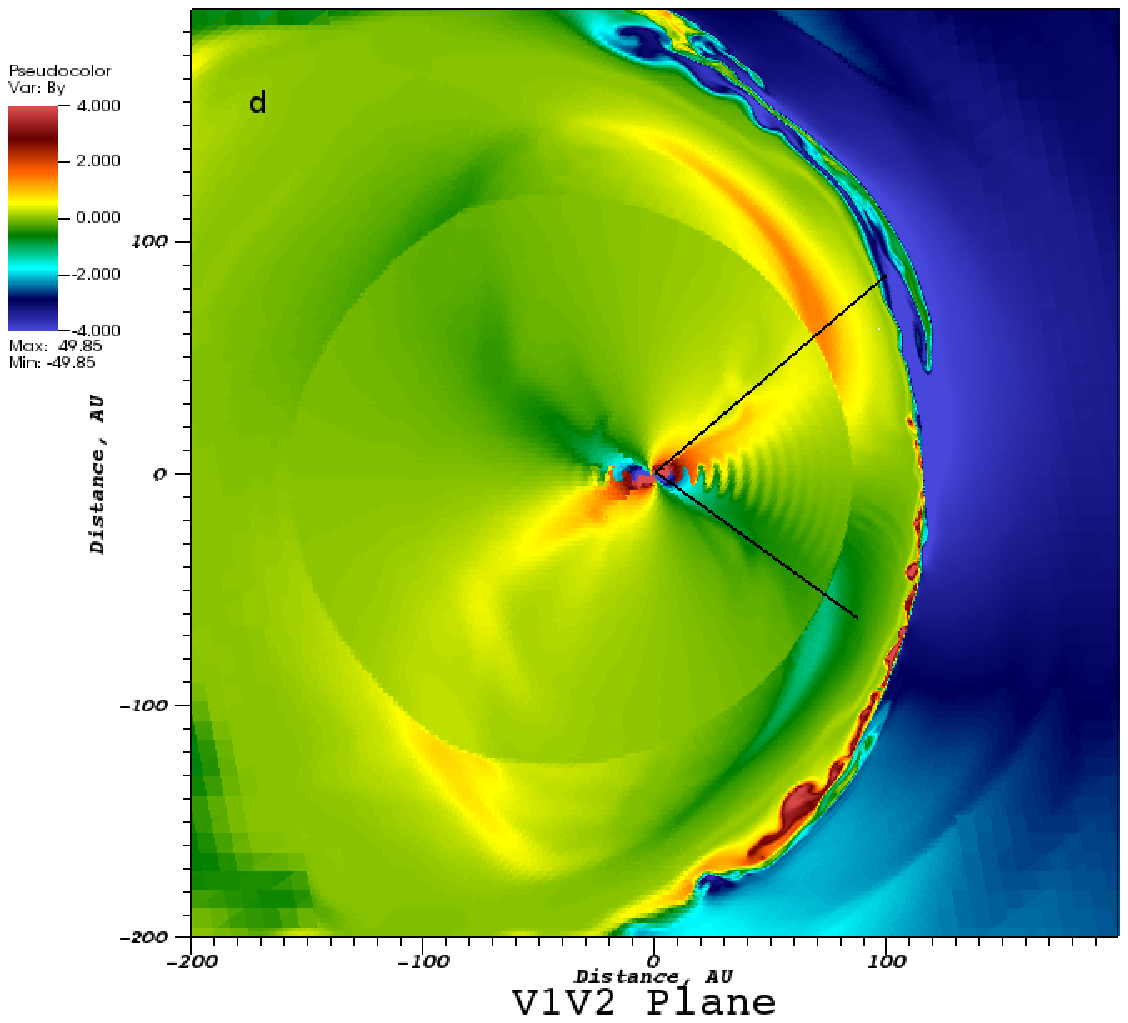}\\
\includegraphics[width=0.45\textwidth]{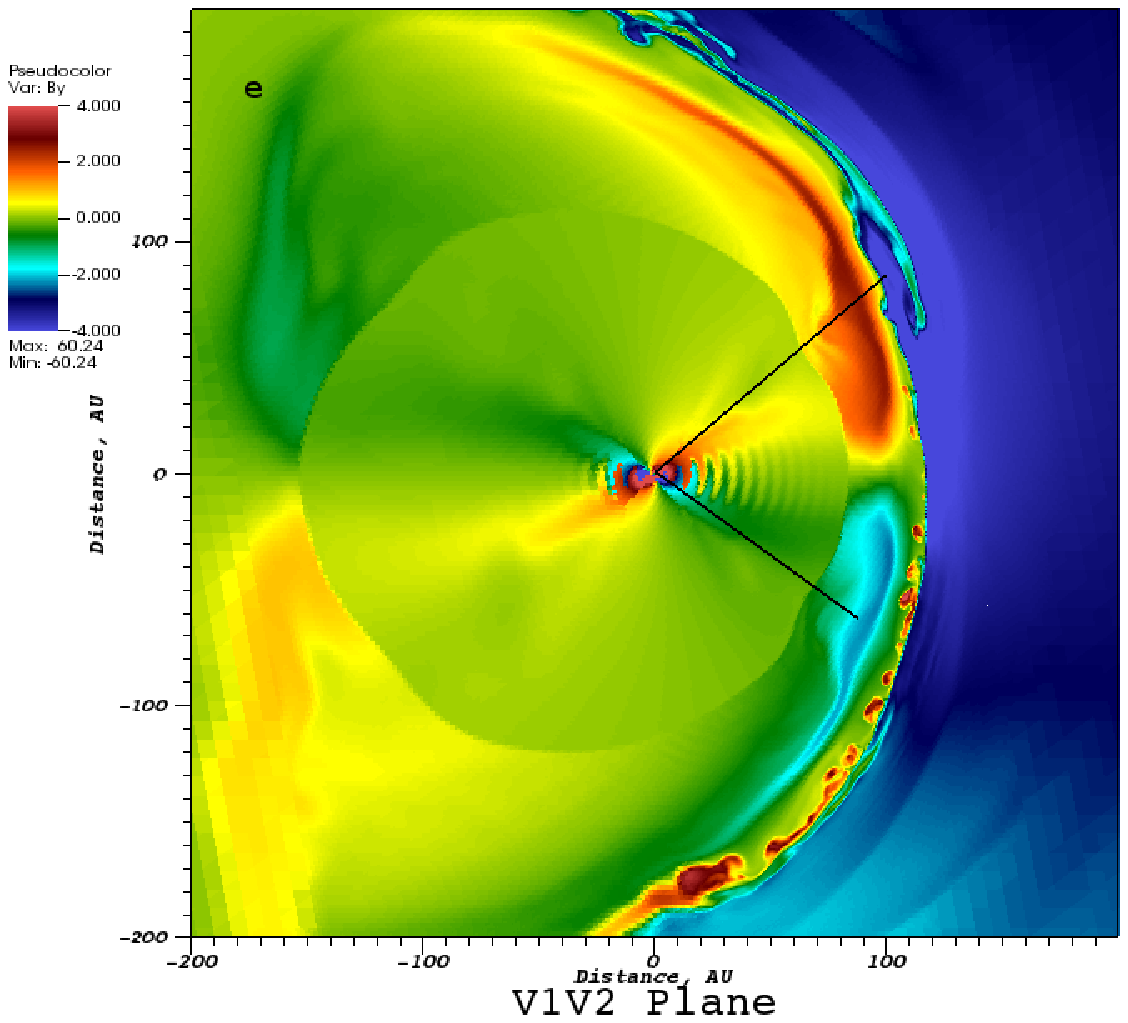}\hspace{5mm}
\includegraphics[width=0.45\textwidth]{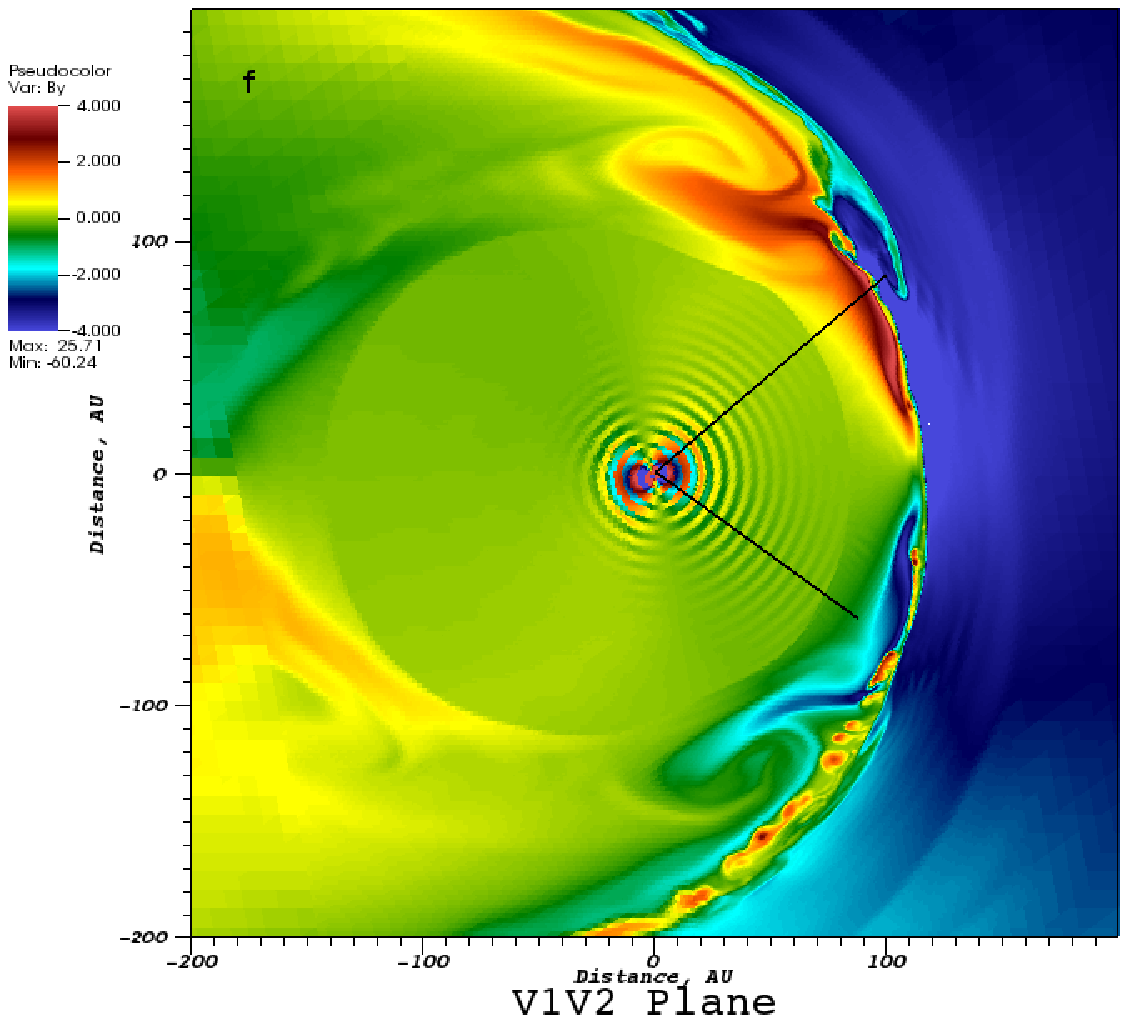}
\caption{The same as in Figure~\ref{fig6} but for $B_y$.}
\label{fig7}
\end{figure*}
\begin{figure*}[p]
\centering
\includegraphics[width=0.45\textwidth]{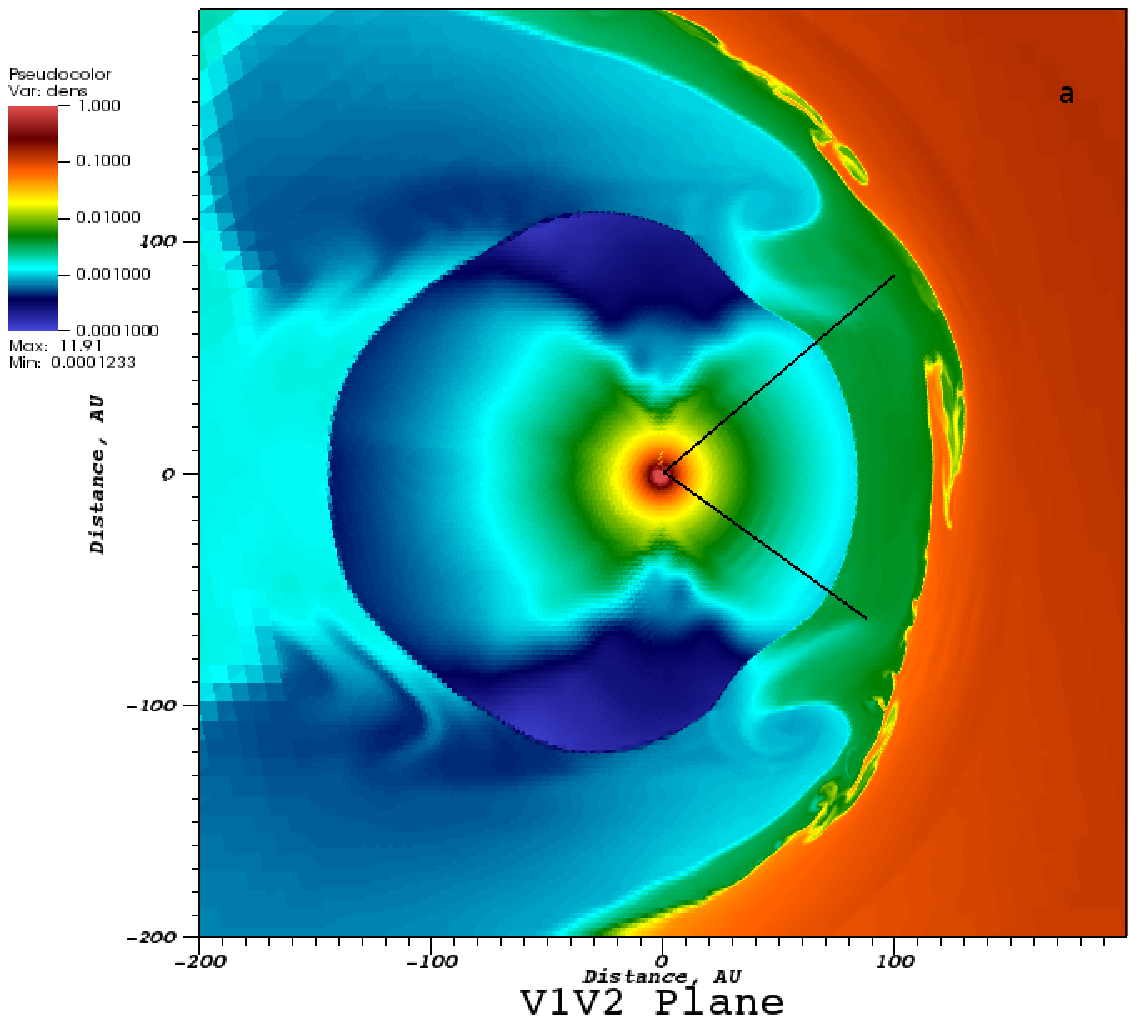}\hspace{5mm}
\includegraphics[width=0.45\textwidth]{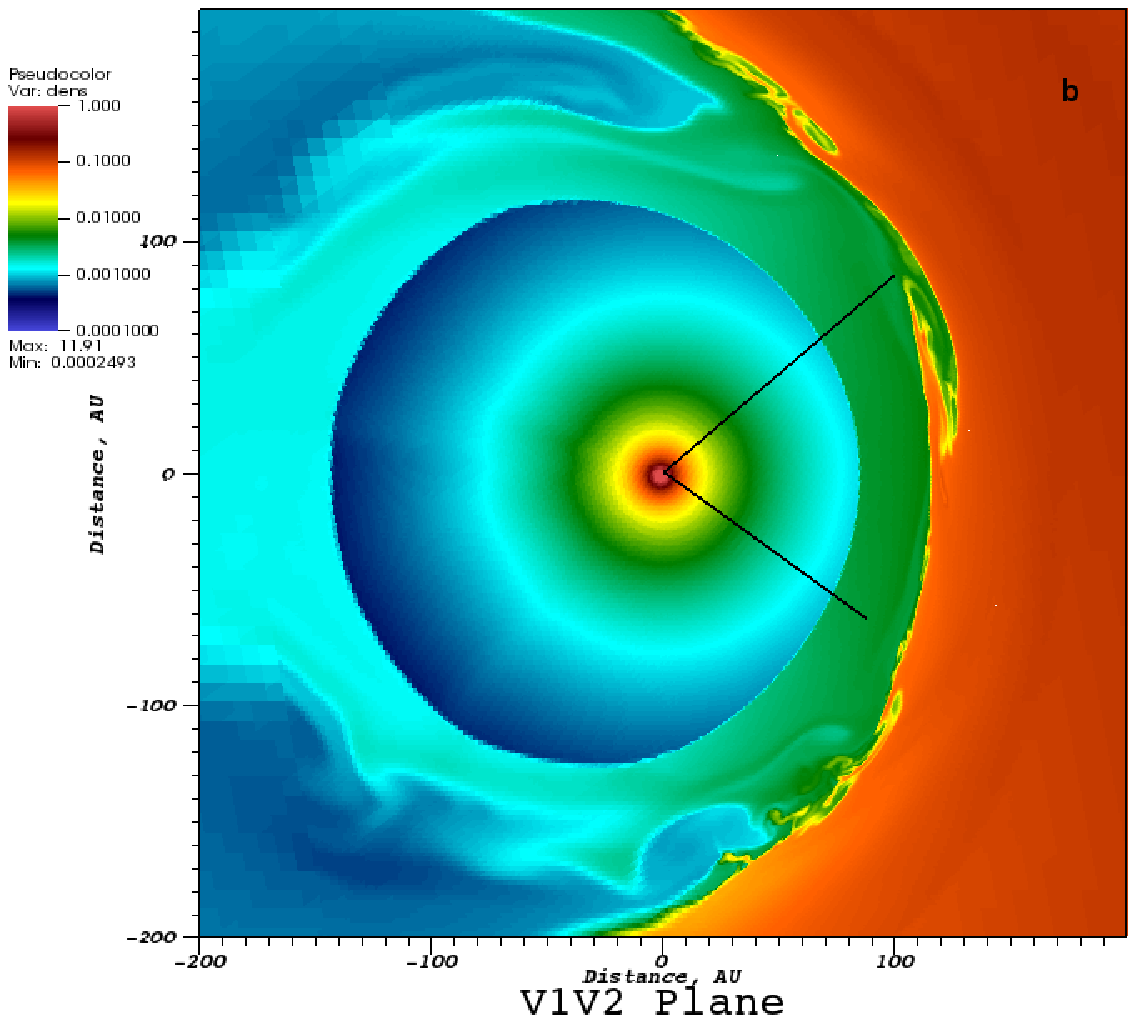}\\
\includegraphics[width=0.45\textwidth]{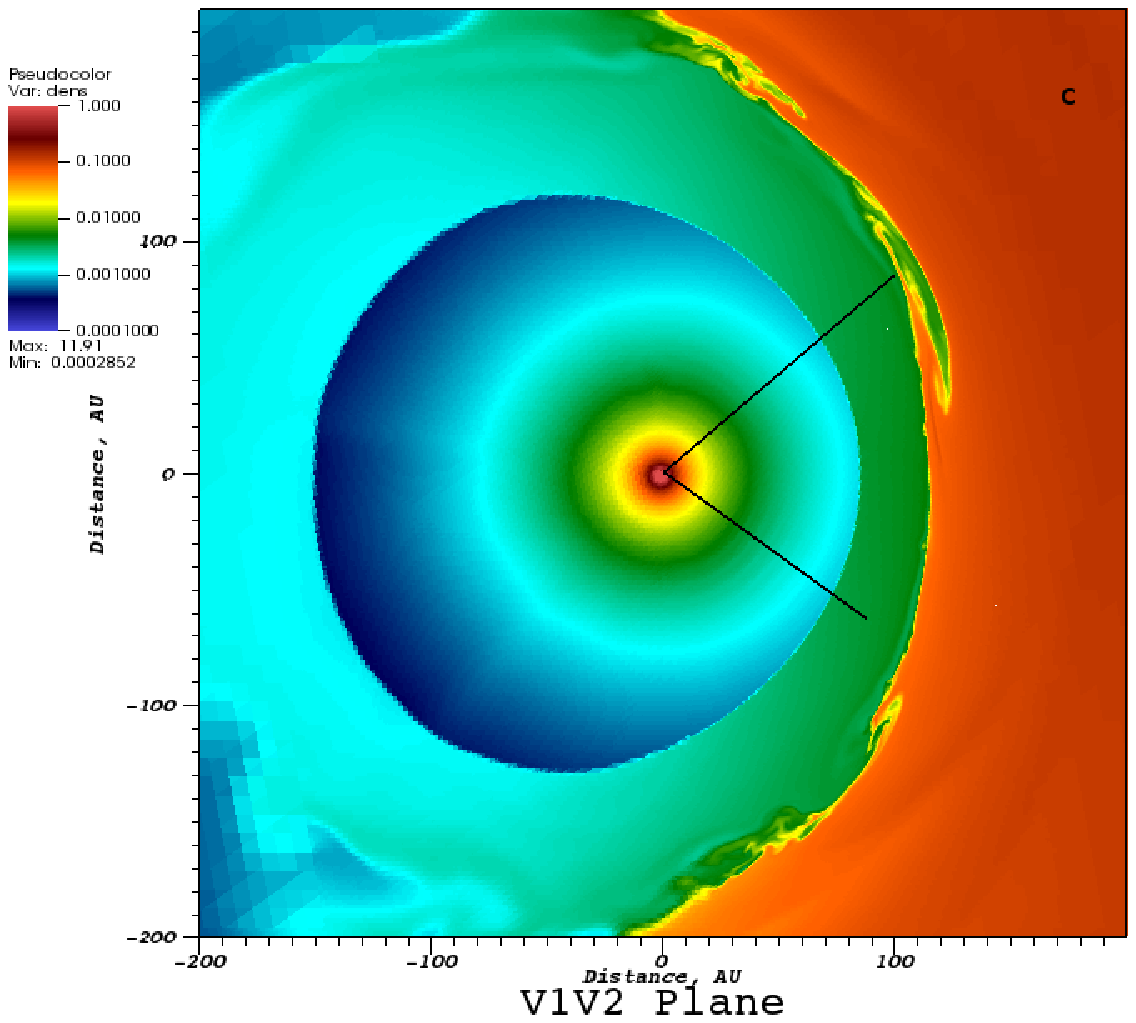}\hspace{5mm}
\includegraphics[width=0.45\textwidth]{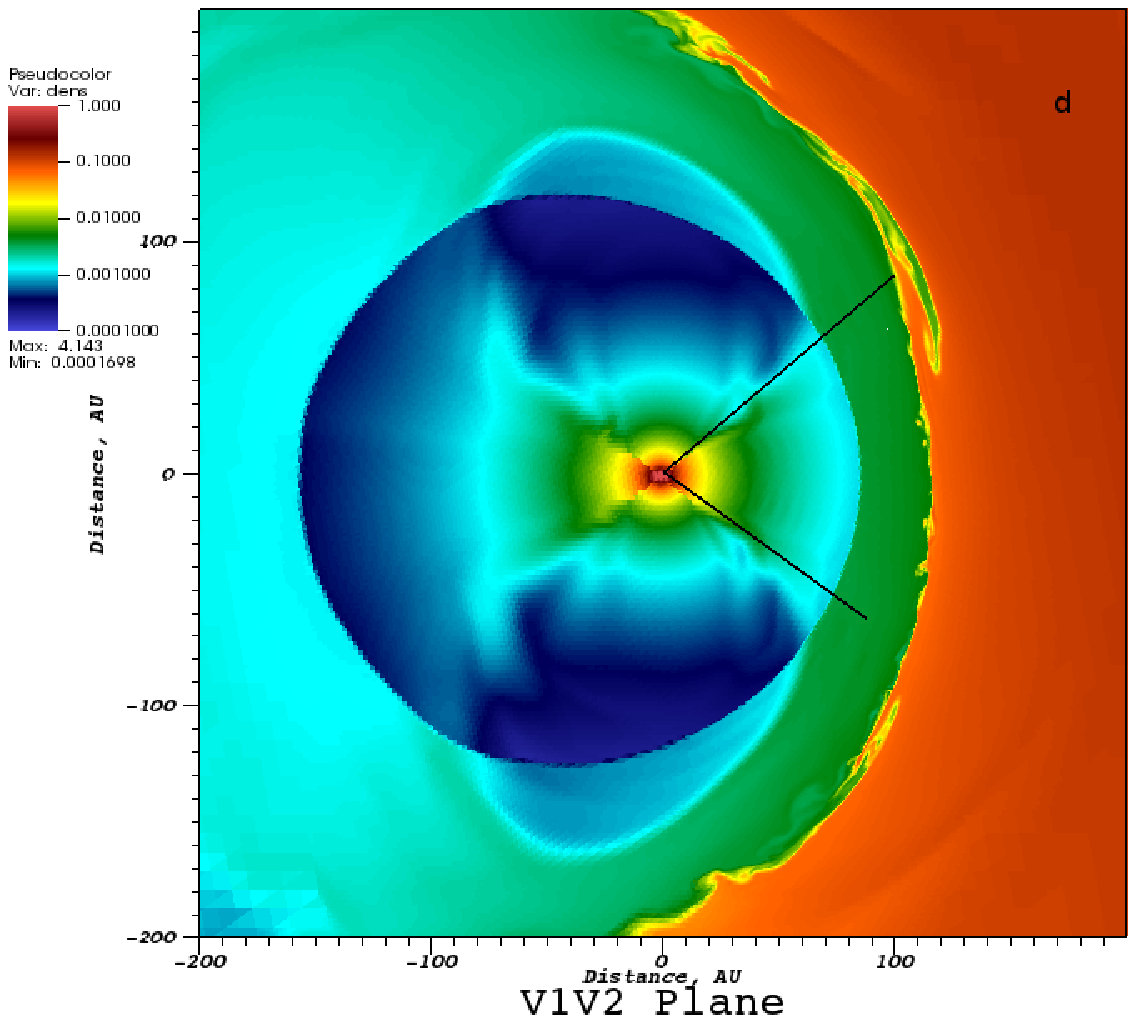}\\
\includegraphics[width=0.45\textwidth]{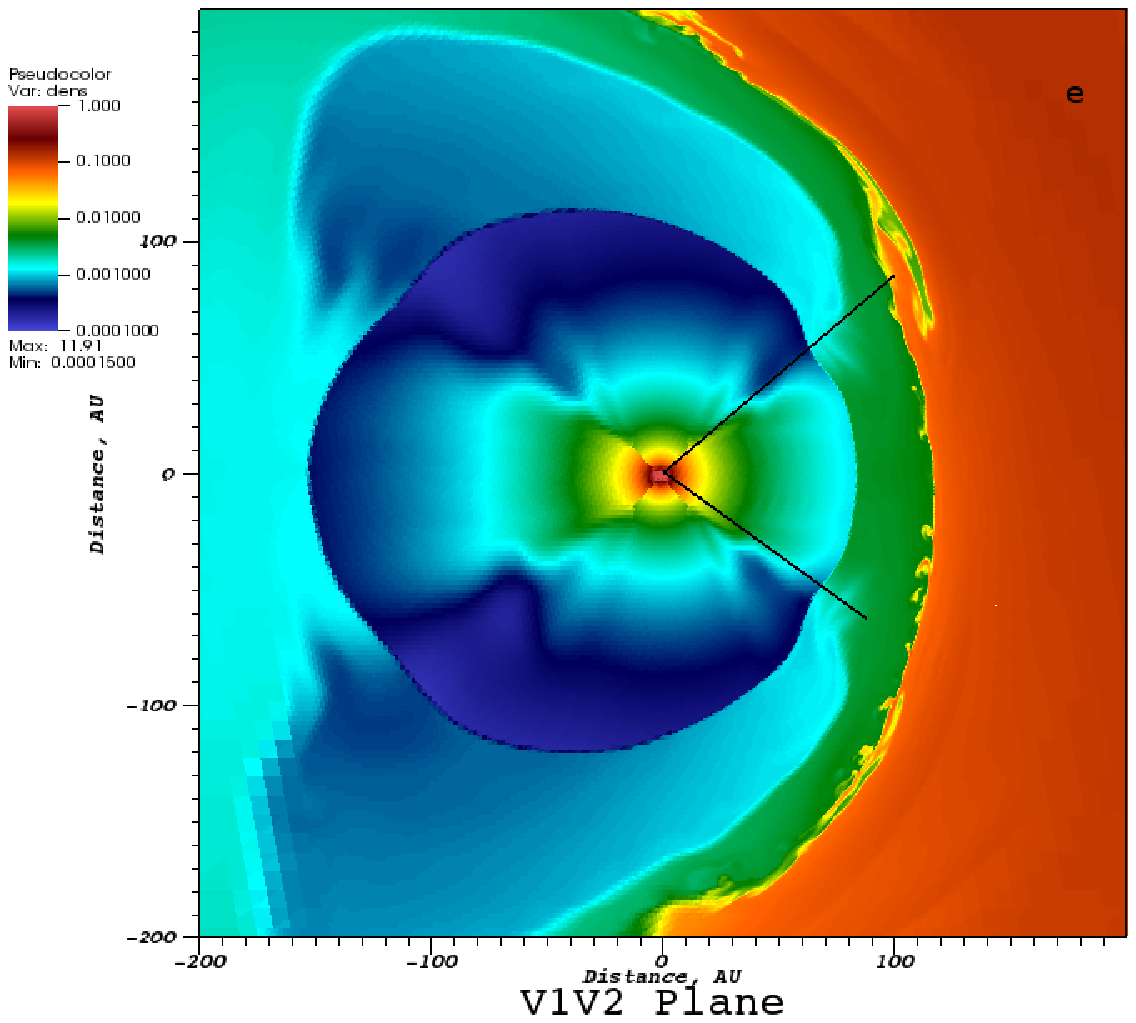}\hspace{5mm}
\includegraphics[width=0.45\textwidth]{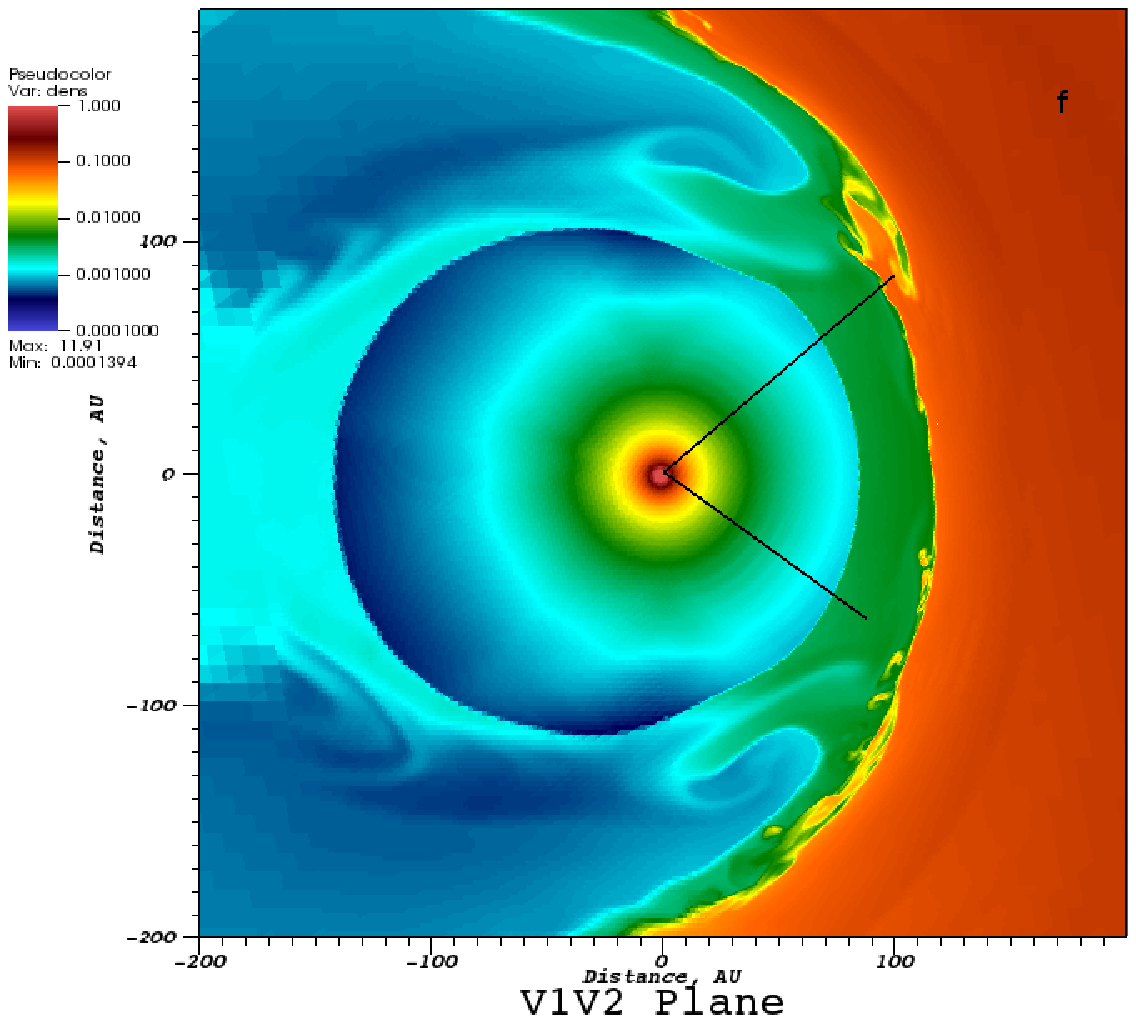}
\caption{The same as in Figure~\ref{fig6} but for plasma density.}
\label{fig8}
\end{figure*}

The RT-instability of the HP results in a penetration of the LISM plasma into the heliosphere. This is clearly seen in Figs.~\ref{fig6}--\ref{fig8}, which show the time evolution of the magnetic field magnitude, $B$, the $y$-component of the magnetic field vector, $B_y$, and plasma density.
The boundary conditions are taken from \citet{Borov14}, but the resolution is higher (0.22~au cubed).
For better understanding of the solution behavior at \textit{Voyager} spacecraft, the cross-cuts are made by the plane defined by the \textit{V1} (in the northern hemisphere) and \textit{V2} (in the southern hemisphere) trajectories.
One can see that the solar cycle creates magnetic barriers of opposite polarity that propagate through the IHS towards the HP.
The HMF polarity in such barriers changes every 11-years.
As a barrier approaches the HP, it becomes exceedingly thinner (see Fig.~\ref{fig7}), creating the possibility of magnetic reconnection across the HP if the orientations of $\mathbf{B}$ become suitable. The inspection of these figures demonstrates that this is especially true for
the southern hemisphere, where \textit{V2} is approaching the HP. As seen in Figs.~\ref{fig7}d--e, magnetic reconnection reveals itself as a
tearing mode (or plasmoid) instability. Similar features are seen in numerical modeling of magnetic reconnection during solar eruptions presented, e.g., in \citet{Pontin}.

It has been shown that plasmoid (tearing-mode) instability of extended current sheets provides a mechanism for fast magnetic reconnection in large-scale systems. Within an MHD framework, the instability has a growth rate that increases with the Lundquist number, while its nonlinear development results in a turbulent reconnection layer and average reconnection rates that are independent of or weakly dependent on resistivity
\citep{Shibata01,Loureiro07,Loureiro13,Bhattacharjee,Uzdensky,Higginson,Higginson17}.
The conditions for such instability are satisfied for high Lundquist numbers, which can be reached by increasing the grid resolution, and large aspect ratios of the reconnecting current layers. While we do not explicitly include resistivity, magnetic diffusion proportional to $\Delta^2$ enters our system due to discretization. Henceforth, both conditions are clearly satisfied in our simulations.

While plasmoid instability ensures that reconnection can remain fast at very large Lundquist numbers, it is by no means the only such mechanism.  Simulations and theoretical considerations demonstrate that in the presence of turbulence large-scale magnetic reconnection can proceed with large, resistivity-independent rates \citep[e.g.,][and references therein]{Lazarian99,Eyink11}.
In fact, \citet{Beresnyak17} argues that the physical reason for the resistivity-independent reconnection rate
is a consequence of turbulence locality, similar to models of reconnection due to ambient turbulence.
Furthermore, under certain conditions, the plasmoid instability can directly transition the system to a kinetic regime where local reconnection rates again become independent of resistivity \citep[see, e.g.,][]{Daughton09,Daughton12,Ji11}. In addition, \citet{Zweibel} show that the presence of neutral atoms may modify the reconnection process. These effects, however, are beyond the scope of this paper and will be addressed elsewhere.

By tracing magnetic field lines that pass through one of the plasmoid regions shown in Fig.~\ref{fig7}, we arrive at another conclusion: the actual magnetic reconnection occurs at a distance of a few AU away from this region. This conclusion has a far-reaching consequence, i.e., magnetic reconnection events have global, macroscopic consequences, which cannot be addressed directly by kinetic simulations because of the length scale limitations intrinsic to them. A similar situations may be observed in solar flares \citep{Liu13}, where the length of a magnetic reconnection sheet is in excess of $10^6$ ion inertial lengths. %Computational boxes of this size are not affordable for kinetic simulations simulations these days.

While more ``reconnection'' is seen in the southern hemisphere and at \textit{V2}, the consequences of the HP instability are stronger at \textit{V1}. Magnetic field distributions in Figures~\ref{fig6}--\ref{fig7} demonstrate the possibility that \textit{V1} could cross
the regions belonging to the SW and LISM consecutively. This means that on the way out of the heliosphere it could be magnetically connected either to the HMF, and observe enhanced fluxes of anomalous cosmic rays (AMRs) and depressed GCR fluxes, or to the ISMF, where ACRs virtually disappear, while the GCR flux increases. This scenario requires that diffusion parallel to the magnetic field should be substantially greater than
perpendicular diffusion. \citet{Luo} show that the  abrupt increase in the GCR flux observed by \textit{V1} when it crossed the HP is possible only if the ratio between the parallel and perpendicular diffusion coefficients exceeds $10^4$. Our simulations provide a plausible explanation of the changes in the ACR and GCR fluxes before \textit{V1} entered the LISM permanently.

To supplement the results shown in Figs.~\ref{fig6}--\ref{fig8}, we add animations of the same quantities to the on-line version of the paper.

\section{Magnetic field dissipation in the IHS}
Issues related to the magnetic field behavior at \textit{V1} and~\textit{V2} are of great importance because both spacecraft provide us with appropriate measurements \citep{B1,B2}. In the idealized simulation considered in the previous section, the angle between the Sun's magnetic and rotation axes is a periodic function of time. The minimum tilt of $8^\circ$ is attained at solar activity minima, whereas the maximum of $90^\circ$ is reached at solar activity maxima, where the magnetic dipole flips from one hemisphere to another. This is, of course, a simplification. \citet{Pogo13b} considered solar cycle effects with the tilt being a function of time from WSO data. As a result,
the magnetic barriers described in the previous section had a layered structure, which was due to local non-monotonicities in the tilt angle
in the vicinity of the spacecraft latitude. Every non-monotonicity of this kind creates an additional current sheet.

Clearly, resolving the sectors of alternating magnetic field polarity in the IHS is impossible, even for a simplified solar cycle.
This is because the sector width is proportional to the SW velocity, provided that the HCS propagates kinematically and exerts no back reaction onto the plasma surrounding it. \citet{Borov11} proposed another approach to track the HCS surface. Other approaches to track the HMF polarity were used in \citet{Czechowski,Alexashov}. In the approach of \citet{Borov11}, the HMF is assumed unipolar and a special, level-set equation is solved to propagate the HCS surface from the inner boundary towards the HP.
Once the HCS surface is known, it is easy to assign proper signs to the HMF vector components at a postprocessing stage. However, this turned out to be impossible even for the level-set approach because the distances to be resolved near the HP become too small for any practically
acceptable grid, so the HCS was accurately resolved only half way from the TS to the HP. In principle, for any chosen grid resolution $\Delta$, this approach fails once $\Delta /T >V$, where $T\approx 25$~days is the period of the Sun's rotation. E.g., for the radial SW velocity component of $\sim 90$~km s$^{-1}$, which is currently observed at \textit{V2} the sector width in the solar equatorial plane is $\sim 1.3$~AU. This means that one would need at least $\Delta \approx 0.13$~AU to resolve the sector structure. Such resolutions are impossible except for
over a very limited region. \textit{Voyager~1}, on the other hand, had been observing negative radial velocity components for two years before it crossed the HP \citep{Decker}. The sector width should be negligible in this case. Moreover, the sector width decreases to zero at the HCS tips \citep[see Fig.~14 in][]{Pogo13b}, which makes attempts to resolve the traditional HCS structure questionable. We call the HCS traditional if the sector structure is entirely due to the Sun's rotation with a fixed period.

\begin{figure}[t]
\centering
\includegraphics[width=\columnwidth]{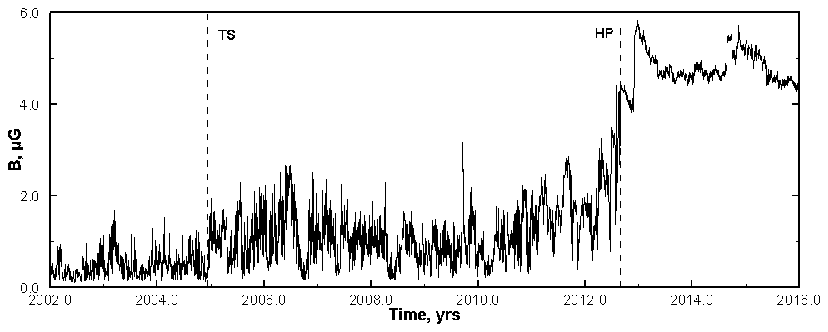}\\
\includegraphics[width=\columnwidth]{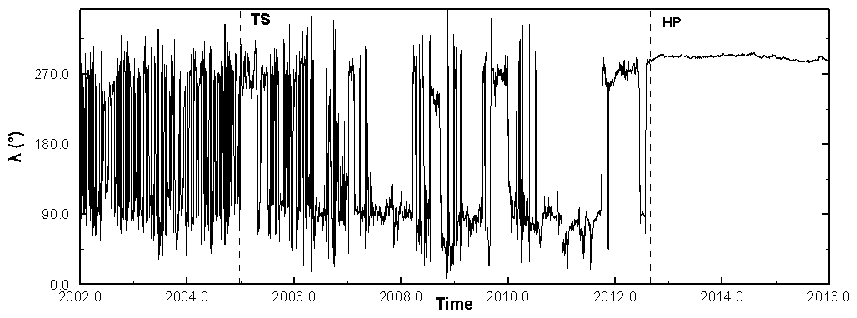}
\caption{Hourly averages of the magnetic field vector magnitude and azimuthal angle along the \textit{V1} trajectory (\textit{Voyager} data courtesy of CohoWeb).}
\label{fig9}
\end{figure}

The intervals between HCS crossings depend on the relative velocity of the SW with respect to the moving spacecraft.
If $V=350$~km s$^{-1}$ in front of the TS and becomes 150~km~s$^{-1}$ behind it, the maximum sector width was about $0.014\times V$~AU, i.e, 4.9 AU in front of the TS and 2.1 AU behind it. The velocities of \textit{V1} and \textit{V2} are approximately 16.6~km s$^{-1}$ and 14.2~km s$^{-1}$, respectively. So \textit{V1} should have been crossing an idealized HCS every 25.5 days before the TS and every 27 days after it, which is not the case (see Fig.~\ref{fig9}).
\textit{Voyager}~2 at the current SW radial velocity of 90 km/s, should cross the sectors at least every 29 days. However, observations show that it is in the unipolar region now. Clearly, the crossing intervals increase, but negligibly, which likely means that an idealized HCS does not exist.

It has been clearly established \citep[see, e.g.,][]{Burlaga94} that a periodic sector structure does not exist beyond 10 or 20 AU. Even at the Earth orbit, the rotating magnetic dipole model produces a two-sector quasi-periodic pattern which is seen only during the declining phase of the solar cycle, when there exist only two dominant polar coronal holes extending towards the equator.
In addition, coronal mass ejections (CMEs), including magnetic clouds, disrupt the sector structure. Evolving coronal holes produce quadruple distortions of the HCS resulting in even more complex sectors. More importantly, however, corotating and transient streams come in different sizes and shapes. They interact with each other and with CMEs to displace and modify any existing structure, while the stream structure itself decays. These are the processes that produce nonperiodic sector structure observed at 30 AU.
It is important, however, that sector boundaries are observed at large distances, at least when they are simple current sheets. This is a likely reason for \citet{John16} to be able to count current sheets associated with sector boundaries.

The fluctuations in magnetic field, density, and velocity components suggest that the HCS is subject to instabilities and is likely torn into pieces as in \citet{Pogo13b}, where it was found that the HCS  does not simply dissipate, but becomes fractured because of the tearing mode instability of the original HCS surface (see the discussion in the previous section).
%Numerical dissipation should be less than some critical value in this case to avoid a complete disappearance of the sector region.
\begin{figure}[t]
\centering
\includegraphics[width=\columnwidth]{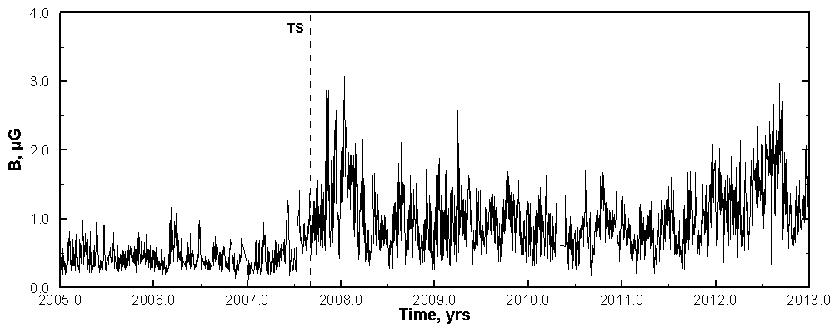}\\
\includegraphics[width=\columnwidth]{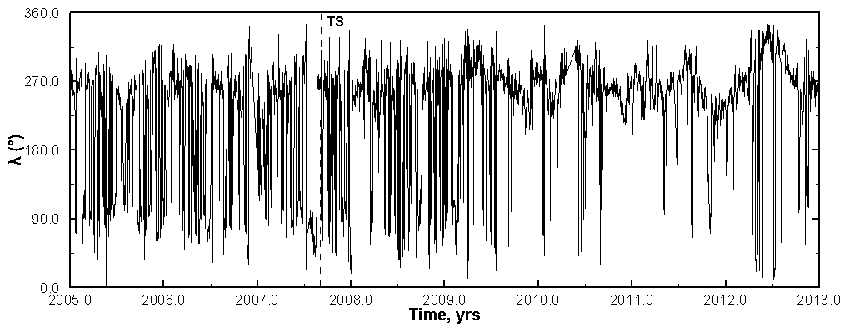}
\caption{Hourly averages of the magnetic field vector magnitude and azimuthal angle along the \textit{V2} trajectory (\textit{Voyager} data courtesy of CohoWeb).}
\label{fig10}
\end{figure}
It should be noticed, however, that the SW is turbulent both inside the TS-bounded region and in the IHS.
Clearly, turbulent fluctuations should increase after the SW crosses the TS, and the HCS may be affected by this turbulence.
Although we know that turbulence affects the HCS, the HP, and magnetic reconnection across them, the question is whether one should
always proceed to kinetic scales to ensure fast reconnection rates, see, e.g., the particle-in-cell simulations of
\citet{Drake10} and the Hall--MHD calculations of \citet{Schreier}. \citet{Lazarian99} identified stochastic wandering of
magnetic field lines as the most critical property of MHD turbulence which permits fast reconnection.
This approach has been successfully validated by \citet{Kowal09,Kowal11,Kowal12,Kowal17}, and \citet{Lazarian11}.
It is clear that ``frozen-in'' magnetic field lines preclude rapid changes in magnetic topology
observed at high conductivities. While microphysical plasma processes demonstrate high reconnection rates
\citep{Che,Daughton11,Moser12}, it is an open question whether such processes can rapidly reconnect astrophysical flux structures much greater in extent than several thousand gyroradii. According to~\citet{Lazarian99,Eyink11,Eyink13}, turbulent
\citet{Richardson} advection brings field lines implosively together from distances far apart to separations of the order of a few gyroradii.
This scenario does not appeal to changes in the microscopic properties of plasma.

This being said, we look at the magnetic field distributions at \textit{V1} (Fig.~\ref{fig9}) and \textit{V2}  (Fig.~\ref{fig10})
spacecraft. It is especially interesting that there is no HCS crossings at \textit{V1} for at least 100 days after crossing the TS.
The HCS is not crossed if a spacecraft moves with the velocity of the ambient SW (17.1 km s$^{-1}$ for \textit{V1} and 15.7 km s$^{-1}$ for \textit{V2}). As shown above, the expected ``nominal'' decrease in the HCS crossing time immediately beyond the TS is small.
This means that either the HMF strength is too small in front of the TS, so that polarity reversals are caused not only by the HCS crossings but also by turbulent fluctuations in the SW plasma, or time-dependent phenomena make the sector structure irregular. Notice
that the sectored region of the SW plasma turns northward in the IHS, so the spacecraft should remain in the polarity-reversal region.

The sector boundaries, if not destroyed by turbulence should be piling up in front of the HP moving inward (at \textit{V1} before it crossed the HP) at $20\ \mathrm{km}\ \mathrm{s}^{-1}$, so the the number of sector crossings should have increased dramatically, but it had not.
Our numerical simulations show that the absence of magnetic field polarity reversals observed by \textit{V1} near the HP may be due to its entering a magnetic barrier. The following few polarity reversals may be caused by the complicated structure of the HP caused by its instability.
\begin{figure*}[p]
\centering
\includegraphics[width=0.45\textwidth]{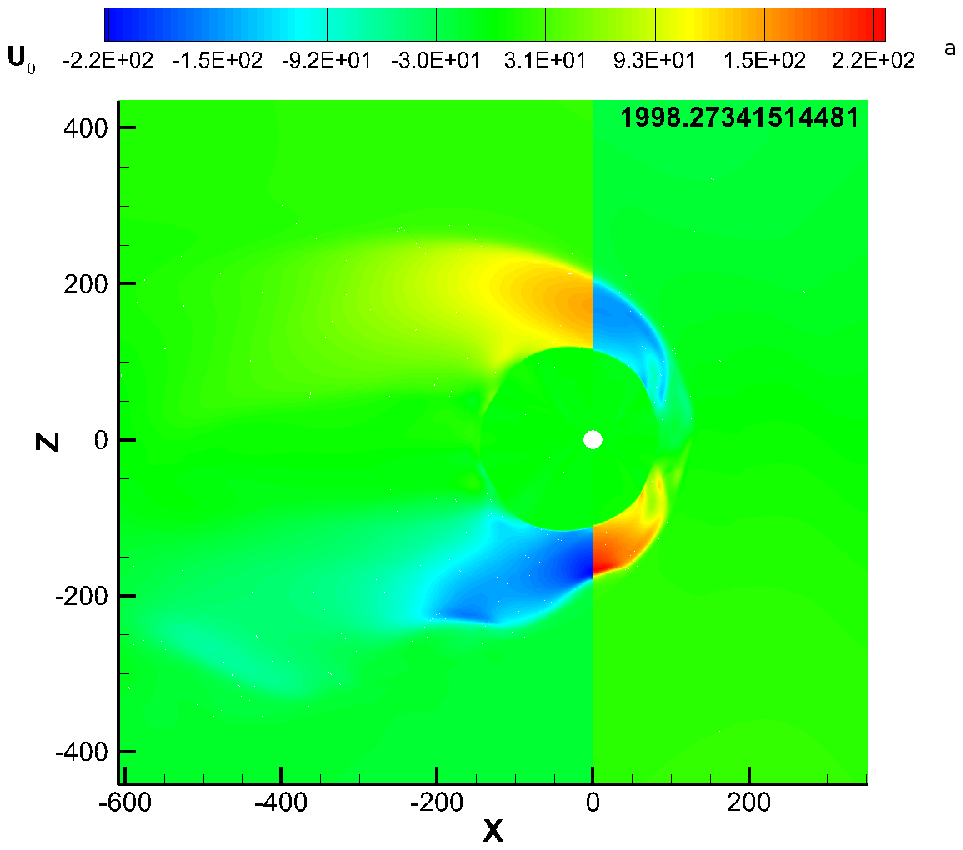}\hspace{5mm}
\includegraphics[width=0.45\textwidth]{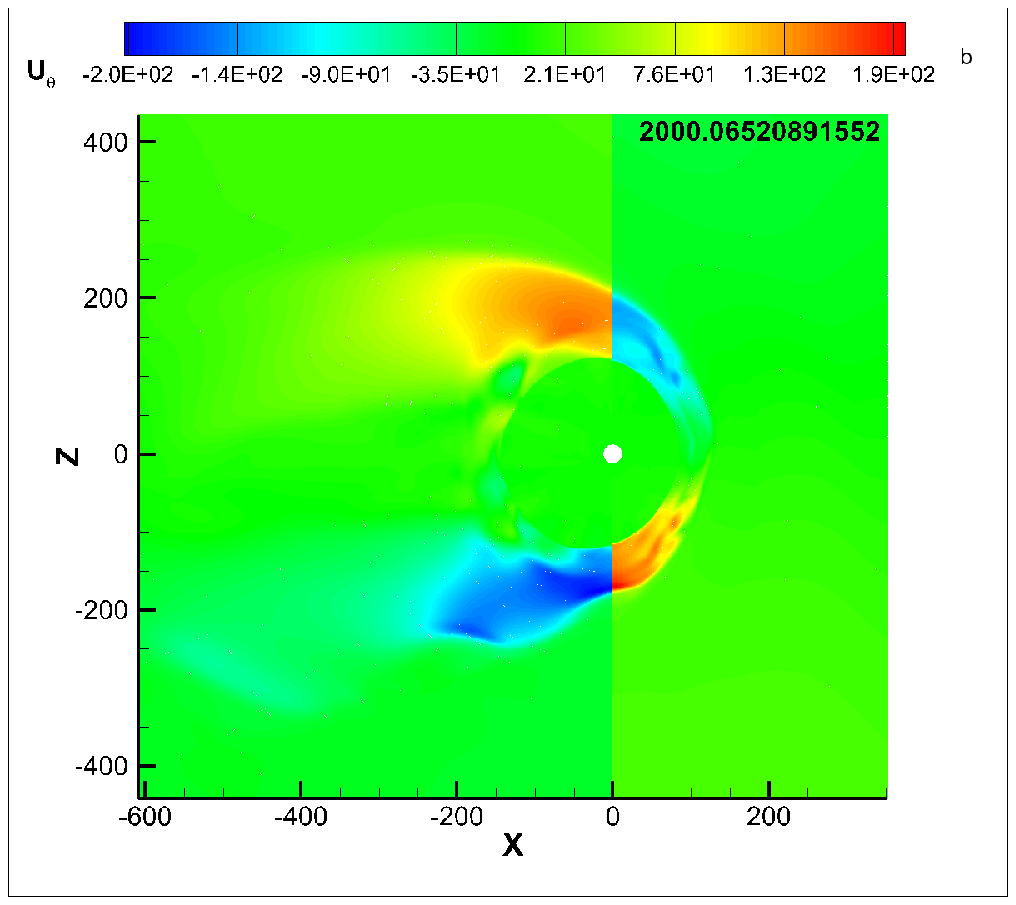}\\
\includegraphics[width=0.45\textwidth]{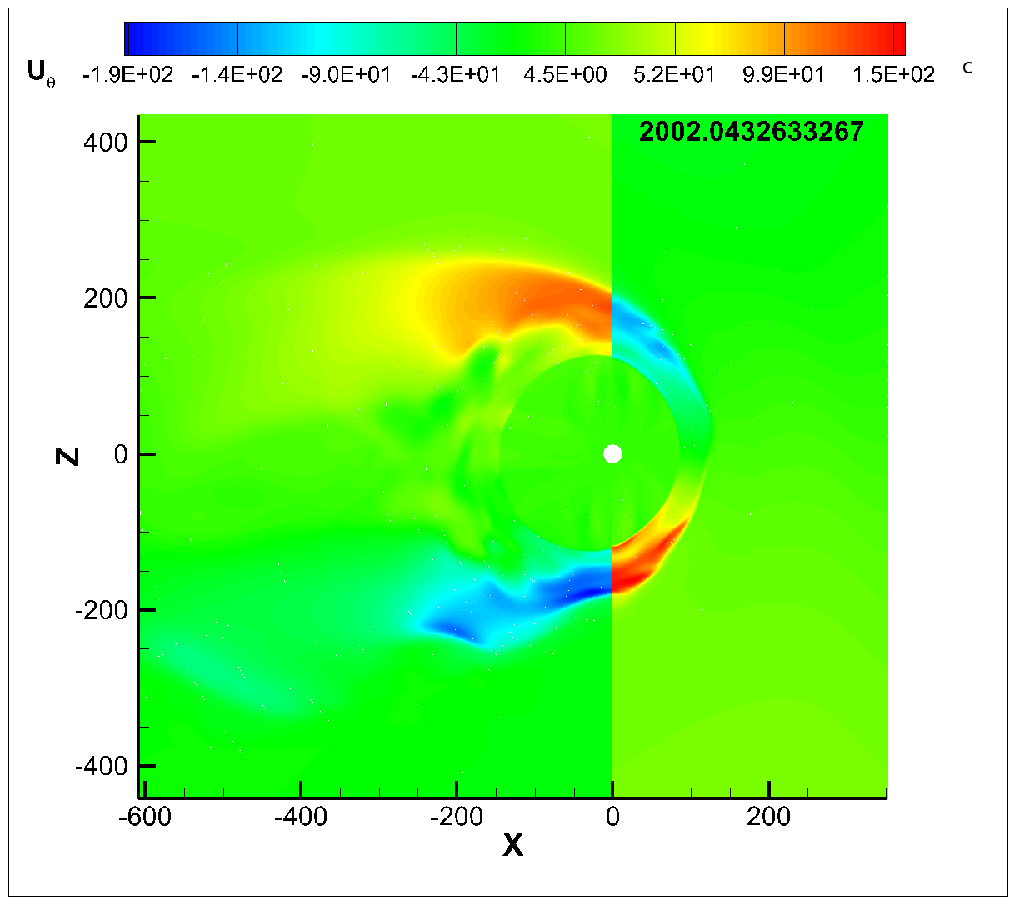}\hspace{5mm}
\includegraphics[width=0.45\textwidth]{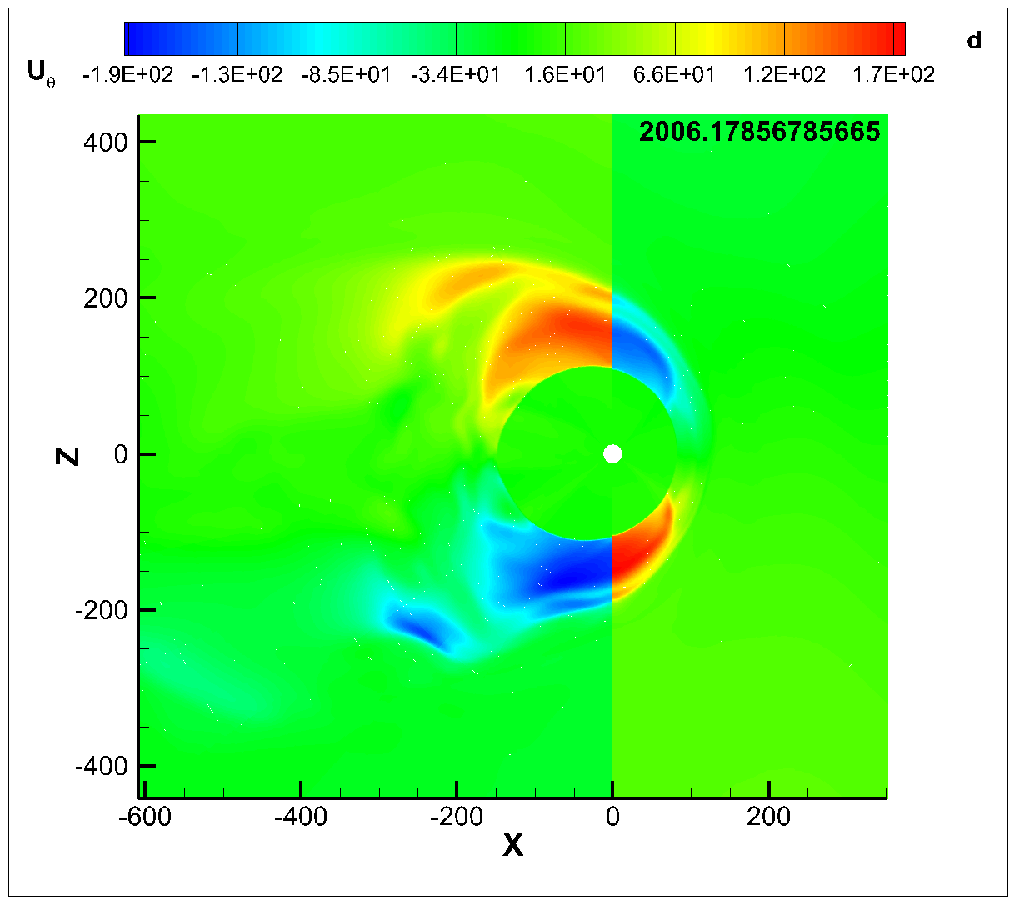}\\
\includegraphics[width=0.45\textwidth]{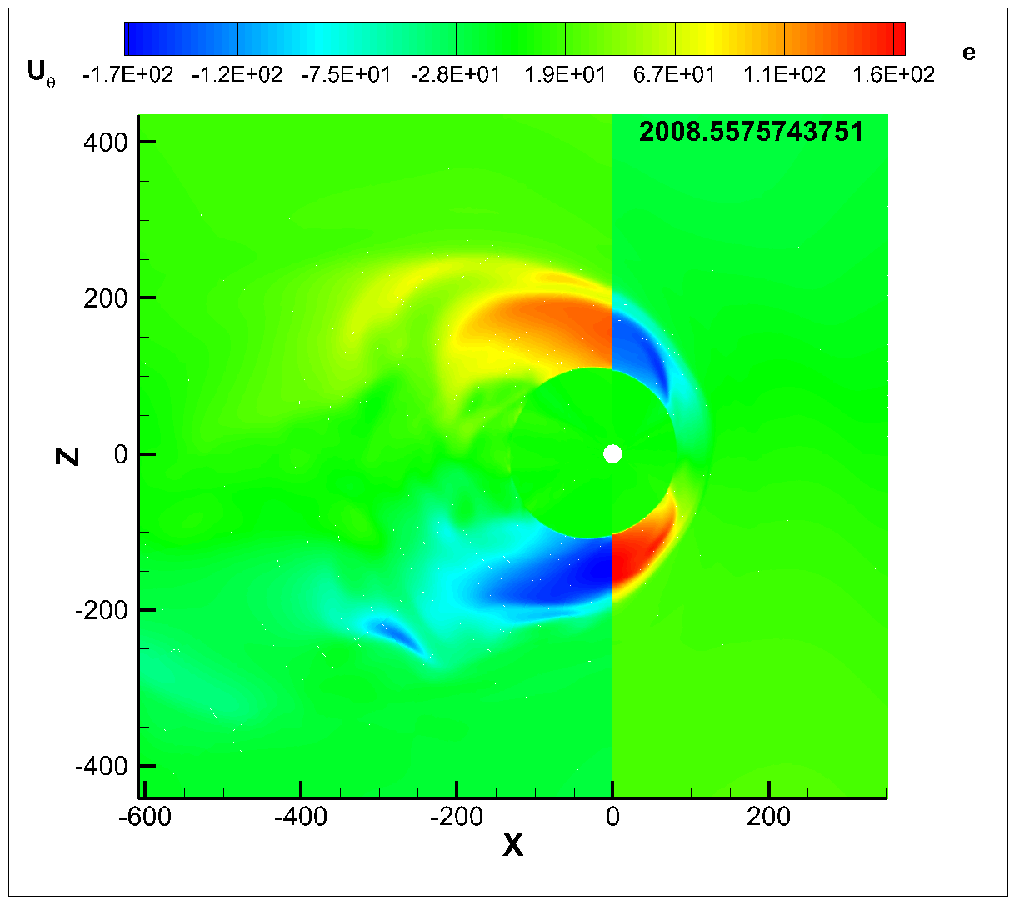}\hspace{5mm}
\includegraphics[width=0.45\textwidth]{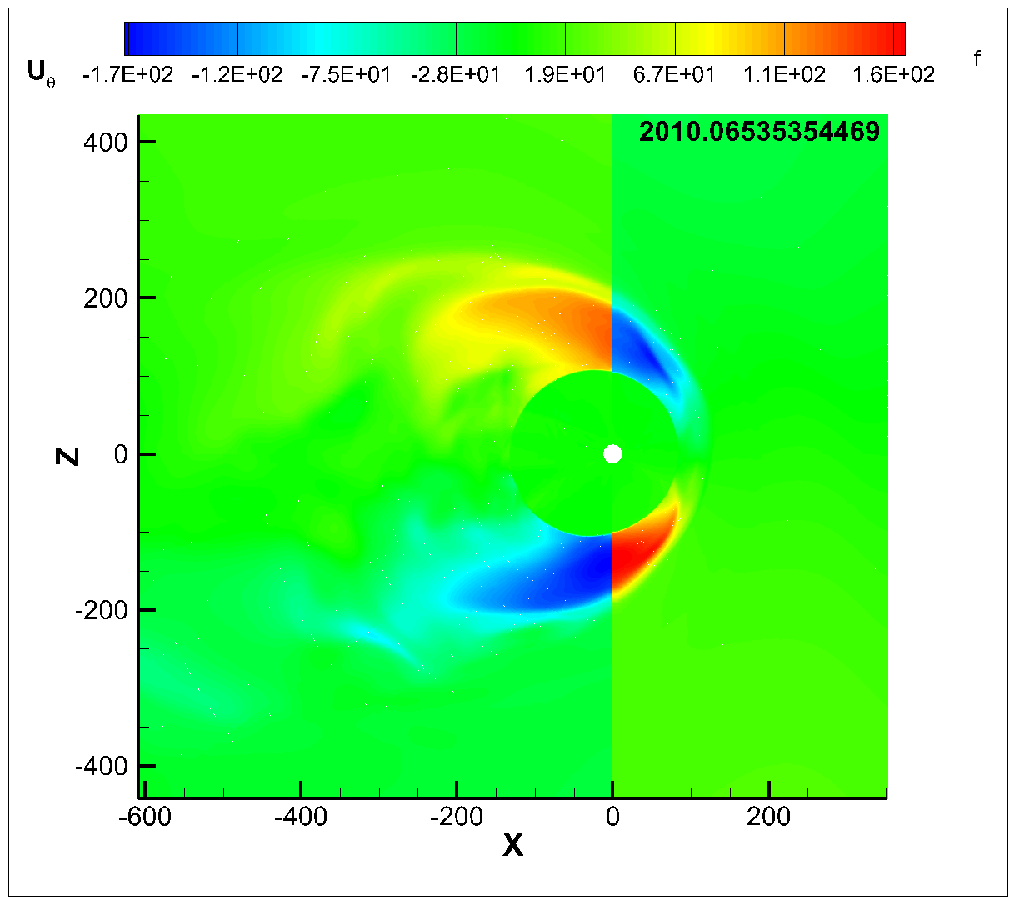}
\caption{Time-dependent distributions of the latitudinal component of the velocity vector in the meridional plane in the simulation of \citet{Pogo13b} demonstrate that this component
strongly depends on latitude.}
\label{fig11}
\end{figure*}

The extended periods with no polarity reversals are also seen along the \textit{V2} trajectory. \citet{John16} have investigated the effect of the magnetic axis tilt on the number of HCS crossings and compared the observed and expected numbers. It has been reported that the number of HCS crossings substantially decreased two years after \textit{V1} and \textit{V2} crossed the TS. However, \textit{V2} might have entered the unipolar region at that time. It was ultimately concluded that there are indications of magnetic field dissipation possibly due to magnetic reconnection across the HCS. However, occasional deviations between the observed and expected HCS crossings are to be expected also for the reasons of stream interaction discussed above and should not be necessarily interpreted as clear evidence for magnetic reconnection.  On the other hand, as shown by \citet{Drake17}, \textit{V2} data reveal that fluctuations in the density and magnetic field strength are anticorrelated in the sectored regions, as expected from their magnetic reconnection modeling, but not in unipolar regions. A possible annihilation of the HMF in such regions may also be an explanation of a sharp reduction in the number of sectors, as seen from the \emph{V1} data.

\citet{John16} assumed that the radial and latitudinal velocity components at the boundary between the unipolar and sectored regions are determined by spacecraft measurements, i.e., are independent of latitude. However, this is not quite true. Figures~\ref{fig11} show that the variations in the latitudinal components can be substantial, which is not surprising because the boundary of the sectored region should propagate from $\sim 8^\circ$ during solar minima to $90^\circ$
at solar maxima in 11 years. Moreover, we remember that a layer of the sectored magnetic field never disappears on the inner side of the HP surface, at least above the equatorial plane. Only its width is a function of time. Thus, a more detailed analysis of observational results
may be required.

\begin{figure}[t]
\centering
\includegraphics[width=\columnwidth]{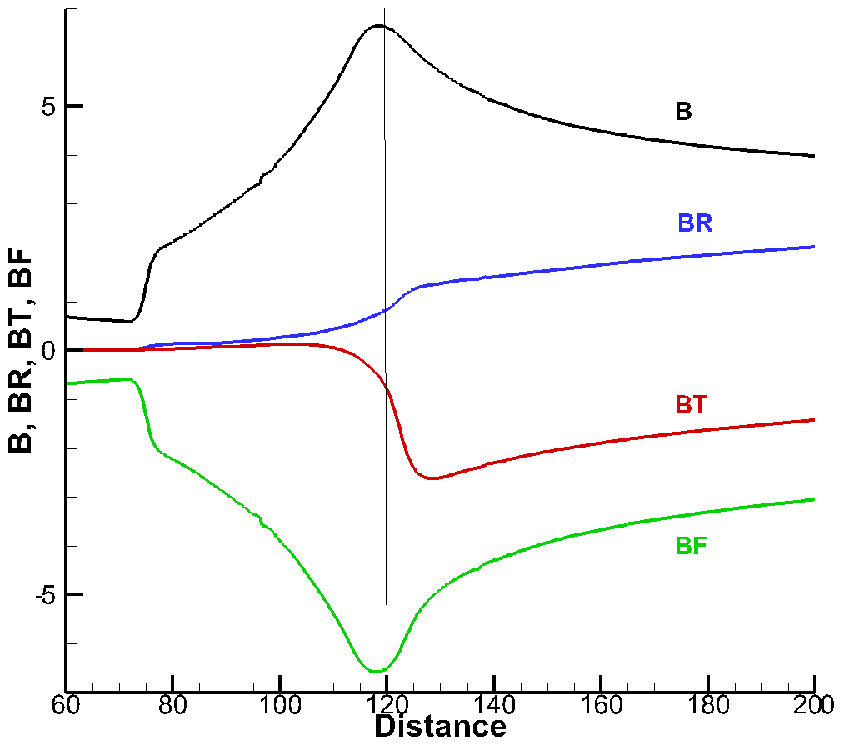}
\caption{The distribution of the magnetic field strength (black line) and its radial (blue line), $\theta$- (red line), and $\phi$-components (green line) along the \textit{V1} trajectory in an MHD-kinetic simulation where the heliospheric magnetic field is unipolar. Presented simulations demonstrate that the assumption of unipolar HMF results in a  considerably overestimation of the magnetic field.}
\label{fig12}
\end{figure}

While the extent to which the HMF dissipates in the IHS remains the subject of investigation, numerical simulations allow us to find out what happens to $\mathbf{B}$ if the HMF is assumed to be unipolar \citep[see the discussions, e.g., in][]{Pogo15,Pogo17}. Figure~\ref{fig12}
shows the magnetic field magnitude, $B$, and the spherical components of $\mathbf{B}$ along the \textit{V1} trajectory for the simulation
with $B_\infty=3\ \mu$G from Table~1. It is clear from this simulation that the calculated magnetic field strength is substantially
overestimated (see Fig.~\ref{fig9}). This behavior of the modeled HMF at small distances beyond the TS was noticed earlier by
\citet{Burlaga09}, but attributed to possible transient effects. On the other hand, based on \textit{Ulysses} measurements, solar-cycle simulations in \citet{Pogo13b}, which take into account the observed variations in the magnetic axis tilt with respect to the rotation axis, although not resolving the HCS, are able to reproduce \textit{V1} observations relatively well on the average. Thus, the possibility of an occasional HMF annihilation in the IHS should not be disregarded.

\section{Conclusions}
In this paper we have addressed a variety of physical phenomena related to \textit{Voyager} observations.
Some of these phenomena have clear physical explanation, whereas additional investigations, both theoretical and numerical, are necessary to
describe the others. The modification of bow shocks by charge exchange between ions and neutral atoms is a well-known phenomenon, frequently referred to not only in the heliospheric bow wave context, but also in astrophysics \citep{Chevalier78,Chevalier80,Blasi12,Morlino12,Morlino13,Morlino16}. Charge exchange cannot modify the Hugoniot relations at shocks
(essentially the conservation laws of mass, momentum, and energy) propagating through plasma, but can substantially change the plasma properties ahead of and behind the shock. As a result, secondary H atoms of the SW origin propagate far upstream into the LISM and decrease the bow shock intensity as compared with ideal MHD flows. The structure of the bow wave in front of the heliopause is of importance for the interpretation of \textit{IBEX} and \textit{Voyager} observations.
In addition, \citet{Nathan14,Ming14}, and \citet{Ming16} demonstrate that it also affects the observed anisotropy of 1--10 TeV cosmic rays.
In the situation relevant to the heliospheric bow shock, the range of possible ISMF strengths, $B_\infty$, is such that the influence of charge exchange becomes dominant, i.e., the contribution of the shock compression to the total density (and magnetic field) enhancement on the HP surface is small. We have shown that any attempt to predict the presence of a shock in the compression region is impossible only by analyzing the properties of the LISM not affected by the presence of the HP. In the absence of such shock, the LISM interaction with the heliosphere produces a rarefaction wave propagating outwards into the LISM.

High-resolution simulations show the presence of a HBL (a region of depressed plasma density and increased magnetic field strength) on the LISM side of the HP. The identification of the internal structure of such layers is a challenge for discontinuity-capturing
numerical methods because of a dramatic change of plasma density across the HP. Discontinuity-fitting methods like that of
\citet{Izmod06} are more suitable for this purpose, unless substantial adaptive mesh refinement is applied.
A drawback of HP-fitting methods in the difficulty to address related physical instabilities.
We demonstrated that the density increase with distance from the heliopause is consistent with the plasma wave frequency observations at \textit{V1} \citep{Gurnett15}. It is of interest that
the boundary layers seen in global simulations are not due to magnetic field effects, since they are also present in simulations without magnetic field
\citep{Bama93,Zank96}. Comparison of multi-fluid and ideal MHD simulations performed by \citet{Pogo05} suggest that the ``depth'' of the boundary layer increases with the LISM neutral H density. On the other hand, the width of the boundary layer seems to be comparable
with the charge exchange mean free path in the LISM (40--50~AU), which means that this boundary layer is somehow related to change exchange.
The relative contribution of the plasma pressure anisotropy on the HBL structure requires further investigation.

While there is little doubt that HP instabilities and magnetic reconnection are intrinsic to the heliospheric interface,
it remains a challenge to relate observational data to simulation results. This is because observations are limited to one point per time,
which makes it difficult to put them into the context of large-scale, 3D phenomena occurring near the HP.
Indeed, the HP instability may result in substantial penetration of the LISM plasma into the SW. This, in turn, creates the possibility for a spacecraft like \textit{V1} to cross the LISM and SW plasmas several times consecutively, which may be a reason for the changes in the observed ACR and GCR flux intensity while \textit{V1} was crossing the HP.
While this scenario still requires confirmation from a direct simulation, it is clear that a 3D, data-driven model of the SW--LISM interaction is necessary to explain the ISMF behavior along the \textit{V1} trajectory. A few attempts to create such model have been presented recently by \citet{Fermo} and \citet{Kim16}. In addition to the HP instability, the simulations presented here predict that \textit{V2} should observe more consequences of magnetic reconnection near the HP than \textit{V1} did. This is apparently the result of the HMF--ISMF coupling at the HP. Further investigation is necessary to understand the physics of plasma wave generation and radio emission observed by \textit{V1} in the OHS. Such investigation should also be data-driven because the ISMF undraping observed by \textit{V1} strongly suggest that time-dependent phenomena are deeply involved in this process. Steady-state MHD-kinetic simulations presented here
\citep[see also][]{Erik16} show that such undraping should be monotone.

The HCS, its behavior in the IHS, and possible influence on the magnetic field and plasma distributions has been one of the most controversial subjects of discussion in the past few years \citep[for a detailed analysis, see][]{Pogo17}. It is clearly impossible to resolve micro-scale phenomena related to the HCS in an ideal MHD model, especially close to the HP. On the other hand, kinetic simulations of magnetic reconnection in the IHS and across the HP are too local to be able to describe the macroscopic effect of this phenomenon. While the HCS is an inherent component of magnetic field distribution in the IHS, numerical simulations allow us to perform a thought experiment where the HCS is excluded by assuming the HMF to be unipolar. As discussed above, one may try to assign the signs to the magnetic field components \emph{post factum}, after a unipolar simulation is finished, provided that we can track the HCS surface as a discontinuity kinematically propagating towards the HP with the SW velocity.
We demonstrated here that this approach is not well justified. If the HMF is assumed unipolar, the simulated magnetic field strength is considerably higher than it was measured by \textit{V1} and \textit{V2}. This is a possible explanation of the discrepancies in the simulations performed with the unipolar and dipolar HMF presented by \citet{Opher15,Pogo15,Pogo16} and further discussed in \citet{Pogo17}.

In summary, our results imply that there is some dissipation of HMF in the IHS. There many reasons for such dissipations: (1) SW turbulence, which is especially enhanced beyond the TS; (2) magnetic reconnections; (3) kinetic and MHD instabilities, etc. A few evidences of such dissipation are discussed in this paper (stochastic destruction of the HCS in the IHS and tearing-mode instability destroying time-dependent magnetic barriers when theirs aspect ratio becomes small). \citet{Pogo13bb} showed that this may result in additional plasma heating and changes in the SW radial velocity component gradients. However, as far as the plasma heating is concerned, it should be clear that its analysis is impossible without proper treatment of PUIs and anomalous cosmic rays from the TS into the IHS. \citet{Pogo16} demonstrated that specifically designed boundary conditions for PUIs at the TS may be able to describe the preferential heating of PUIs as compared with thermal SW ions \citep{Zank10}. Such boundary conditions are of kinetic nature and therefore cannot be derived from any continuum mechanics approach. Approaches which are not based on the conservation-law principles and involving straightforward calculations of the PUI pressure derivatives across the TS \citep[e.g.,][]{Usmanov16} are mathematically flawed. \citet{Fahr} (see also references therein) proposed a number of theoretical approaches to derive the above-mentioned boundary conditions. Local particle simulations may also serve as tools to help derive the boundary conditions that would be able to reproduce the ion acceleration at any point of the TS. The ion distribution function in the shock vicinity is highly anisotropic, which makes continuum mechanics approaches not applicable. We note also a recently proposed generalized system of such equations that take into account dissipative affects and heat flux terms \citep{2014ApJ...797...87Z}, but not taking into account possible reflection of PUIs and their further acceleration at the TS.

\acknowledgments
This work was supported NASA grants NNX14AJ53G, NNX14AF43G, NNX15AN72G, and NNX16AG83G.
The work done in the University of Iowa was supported by NASA through Contract 1279980 with the Jet Propulsion Laboratory.
This work was also partially supported by the \emph{IBEX} mission as a part of NASA's Explorer program.
We acknowledge NSF PRAC award ACI-1144120 and related computer resources from the Blue
Waters sustained-petascale computing project. Supercomputer time allocations were also provided on SGI Pleiades by NASA
High-End Computing Program award SMD-16-7570 and Stampede by NSF XSEDE project MCA07S033.

The authors are grateful to G.~P.~Zank for stimulating discussions.

\end{document}